# Disaggregation for Energy Efficient Fog in Future 6G Networks

Opeyemi O. Ajibola, Taisir E. H. El-Gorashi, and Jaafar M. H. Elmirghani, *Fellow, IEEE*

***Abstract*—We study the benefits of adopting server disaggregation in the fog computing tier by evaluating energy efficient placement of interactive apps in a future fog 6G network. Using a mixed integer linear programming (MILP) model, we compare the adoption of traditional server (TS) and disaggregated server (DS) architectures in a fog network comprising of selected fog computing sites in the metro and access networks. Relative to the use of TSs, our results show that the adoption of DS improves the energy efficiency of the fog network and enables up to 18% reduction in total fog computing power consumption. More instances of interactive fog apps are provisioned in a fog network that is implemented over a network topology with high delay penalty. This ensures that minimal delay is experienced by distributed users. Our result also shows that the proximity of fog computing sites such as metro-central offices and radio cell sites to geo-distributed users of interactive fog applications make them important edge locations for provisioning moderately delay sensitive fog apps. However, fog applications with more stringent delay thresholds require in situ processing at directly connected radio cell sites or at the location of the requesting users. Finally, we propose a heuristic for energy efficient and delay aware placement of interactive fog apps in a fog network which replicates the trends observed during comprehensive analysis of the exact results obtained by solving the MILP model formulated in this paper. Our results and proposed MILP and heuristic provide a good reference and tool for fog network design and deployment.**

## I. Introduction

CLOUD COMPUTING became an integral part of the global society over the last two decades because it enabled improved capital and operational efficiencies relative to the traditional distributed computing architecture. Adoption of cloud computing in its different service offerings such as Infrastructure as a Service (IaaS), Platform as a Service (PaaS) and Software as a Service (SaaS) spans across a variety of personal, enterprise and public applications. In 2020, it is predicted that 83% of enterprise workloads will be in the cloud. Some of the key drivers of this trend include digital transformation, artificial intelligence and machine learning and Internet of Things (IoT) [1]. By 2022, it is estimated that the global spending on IoT will reach $1.2 trillion [2]. Furthermore, Cisco predicts that about 500 billion smart devices will be connected by 2030 [3].

This growing uptake of the cloud computing and IoT paradigms is expected to enable a new range of emerging and future Internet applications at the edge of telecommunication networks. However, the quasi-distributed or semi-centralized architecture of the traditional cloud computing paradigm is a major inhibiting factor to the emergence of some of these applications which may require real-time or near-real-time computation. In addition, the volume, velocity, and variety of data generated by geo-distributed IoT-devices and end-users of these emerging Cloud-IoT applications combined with traditional network traffic is expected to overwhelm network infrastructures. This will also significantly increase the cost of owning and operating these networks and may increase their carbon footprint. Ultimately, the centralized cloud computing architecture will degrade the performance of some future applications due to increased network congestion. In other cases, the centralized cloud computing architecture will out rightly prevent the emergence of some applications which are infeasible under the centralized cloud computing architecture. Hence, the concept of edge/fog computing [4]–[6] has been proposed in recent times to address some of these challenges.

The fog computing paradigm extends cloud computing to the network edge to support emerging and future internet applications and services. This is enabled via the introduction of a new intermediate computation tier called the fog computing tier. The fog computing tier is located between the centralized cloud computing tier and geo-distributed IoT-devices and end-users of emerging applications at the network edge. The fog tier comprises of heterogeneous devices and nodes called fog nodes such as edge routers, access points, specialized servers, and a range of endpoints including connected vehicles, surveillance cameras and mobile phones. These heterogeneous fog nodes/devices adopt heterogeneous network infrastructures (both wireless and wired) for connectivity. The geo-distributed heterogeneous nodes may also be orchestrated collectively as a fog federation within a given area to enable a fog as a service (FaaS) or fog Infrastructure as a serive (IaaS, FIaaS) business

This work was supported in part by the Engineering and Physical Sciences Research Council (EPSRC), in part by INTelligent Energy aware NETworks (INTERNET) under Grant EP/H040536/1, in part by SwiTching And tRansmission (STAR) under Grant EP/K016873/1, and in part by Terabit Bidirectional Multi-user Optical Wireless System (TOWS) project under Grant EP/S016570/1. All data is provided in the results section of this paper. The first author would like to acknowledge his PhD scholarship awarded by the Petroleum Technology Trust Fund (PTDF), Nigeria.

The authors are with the School of Electronic and Electrical Engineering, University of Leeds, Leeds, LS2 9JT, U.K. (e-mail: el14oa@leeds.ac.uk; t.e.h.elgorashi@leeds.ac.uk; j.m.h.elmirghani@leeds.ac.uk).

model [7]–[9]. The goals of the fog computing paradigm include: minimization of response time for real-time and mission critical applications (such as vehicle to vehicle communication, intelligent processing/analytics, wireless sensors and actuation networks and game streaming) via in-situ computation [10], [11]; the reduction of cloud computing destined workloads; and the reduction of total end to end traffic in networks [10]. Like the cloud computing tier, fog nodes possess compute, storage, and networking capabilities. Virtualization might also be supported in fog nodes to ensure efficient use of the computation capacity as obtainable in the cloud computing tier [7]. However, fog nodes are unable to enjoy other benefits obtainable in centralized cloud computing infrastructures. For example, the ability to strategically position massive computational resources at little or no opportunity cost is limited in the fog computing tier. Furthermore, the fog computing nodes cannot fully maximize the benefits of using commodity hardware as obtainable in hyper-scale cloud datacenters (DCs) because they are relatively smaller in size. In recent times, hyper-scale DC infrastructure providers such as Facebook and Microsoft have also begun to explore the use of the server resource disaggregation concept as a tool to further improve on DC infrastructure overall efficiency [12].

Server resource disaggregation proposes the physical or logical separation of traditional server (TS) intrinsic resources into pools of homogeneous resources, which can be composed, decomposed, and recomposed on-demand over high bandwidth and low latency networks, to support applications. This concept addresses the limitations associated with TSs such as poor resource modularity and lifecycle management, the need for purpose-built servers in computing clusters, high power consumption and capital expenditure resulting from inefficient resource utilization [13]–[16]. Adoption of such a concept in the fog computing tier could improve fog computing efficiency to approach efficiencies that are traditionally attributed to the cloud computing tier. In this paper, we explore the gains that resource disaggregation can enable in fog networks relative to TS architecture and study the impact of fog computing on metro networks. In [17], we conducted an initial study to demonstrate some potential benefits of adopting disaggregated servers (DSs) in the fog computing era, concluding that the network bottlenecks may inhibit optimal performance and users experience may suffer due to the resulting increased hop counts between users and fog app instances. This paper extends our initial work as follows; (i) it provides for the first time the complete MILP model, which has been extended by considering queuing delay in the network topology; (ii) it considers emerging fog apps with varying delay sensitivity; (iii) it investigates the impact of user distribution on the fog network; and finally, (iv) it develops a fast and scalable policy (heuristic) which mimics the MILP model for practical deployment in large scenarios, and to verify the MILP.

The rest of this paper is organized as follows. We review the concepts of fog nodes federation and resource disaggregation and related literatures in Section II. Section III presents the system setup and the MILP model formulated to investigate the placement of applications in fog networks. Section IV gives the evaluation scenarios along with numerical results and discusses the insights obtained by solving the MILP model. Section V presents a heuristic that mimics the results and insights obtained from the MILP model. Finally, the paper is concluded in Section VI along with a brief discussion of future work and future directions.

## II. Fog Networks and Disaggregation

### *A. Fog Networks*

The fog computing layer is an intelligent intermediate layer between centralized cloud computing servers and geo-distributed connected devices and end-users. This layer provides distributed computing infrastructure (computation, storage, and networking resources) at network edges (i.e. metro and access networks) to support connected things and end-users. Hence, the fog computing layer complements the centralized hyper-scale cloud computing infrastructure by extending cloud-like services closer to end-users and connected things for improved performance and to support new classes of applications. The fog computing layer reduces application response time, the volume of telecom network traffic and the workload on public cloud infrastructure. Consequently, the fog computing layer can also enable reductions in computing and network infrastructures [18]–[20]. The fog computing paradigm also enhances the performance of some existing and emerging applications such as IoT, content delivery network, artificial intelligence, and data analytics. Furthermore, the fog computing layer enables a suitable environment for future internet applications and services. Additionally, it also provides better support for mobile connected-devices and end-users, better context and location awareness and improved security relative to the situation when the fog layer is absent [7], [11].

In recent times, notable wired communication equipment vendors have included extra computation capacity in network routers/switches to support the hosting of non-network related functions and application. For example, Cisco's catalyst 9000 series switches support application hosting capabilities to host fog applications at the edge of the network [21]. In the wireless communication domain, the concept of mobile edge computing (i.e. a use case of fog computing) is expected to feature in the 5G mobile network infrastructure which is in the early deployment phase. Furthermore, it is also predicted that this concept of mobile edge computing will become more prominent when the deployment of 6G mobile network infrastructure begins. Machine Learning (ML) and Artificial intelligence (AI) are predicted to be key features of 6G network infrastructure [22]. Hence, edge computation is required to provide seamless access to ML / AI capabilities at the network edge in the 6G era.

Traditionally, connected devices and end-users in the "things tier" at the farthest edge of the network interact directly with quasi-distributed cloud computing tier via communication networks (i.e. access, metro and core networks). The introduction of the fog computing tier expands the traditional architecture of the cloud of things continuum [7], [9] by one tier. The fog computing paradigm provides geo-distributed low-medium sized computation and storage capacity near end-users and connected devices while cloud computing provides centralized hyper-scale computation and storage capacity, but with high service delay. The traditional definition of fog computing classifies any device with compute, storage and network connectivity as a fog node [10]. However, some locations and devices maybe more optimal than others because of factors such as energy efficiency, resource capacity, node availability, resource reusability and utilization efficiency. For example, the things tier, which generally comprises of sensors and actuators, is often characterized by small installed computation capacity and limited network connectivity. Hence, resource availability and utilization efficiency at the things tier is often limited. Addressing these limitations by increasing device form-factor and expanding network connectivity in the things tier might be an overkill for the purpose. On the other hand, increasing the number of end-devices to scale computation capacity at the things tier will further increase the total cost of ownership (TCO), power consumption and carbon footprint associated with the tier. Hence, reducing computation capacity at the extreme edge of the network to its optimal limits based on application requirements is important for sustainable growth in the 6G fog networking and computing era.

At the same time, the placement of fog computational capacity in traditional network nodes such as central offices (COs) and radio cell sites (CSs), where it can be easily accessed by multiple connected devices and end-users, is another sustainable strategy that could be adopted. Such strategy also provides effective support for expected high mobility of end-users and connected devices in the fog networking and computing era [11], [23], [24]. Additionally, sharing of existing medium-sized computation capacities at edge locations such as enterprise offices (EOs) and public locations can also enhance sustainability in the fog computing era. The proliferation of geo-distributed micro-DCs such as the hyper-converged infrastructure (HCI) nodes and the use of commodity hardware to support network function virtualization (NFV) at the edge of the network also supports the business case for centralized coordination and scaling of shared fog computing capacity. For instance, the authors of [25] demonstrated the importance of NFV as a key technology for energy efficiency in 5G networks. Such technological advantage can be leveraged to improve the energy efficiency of the general fog computing layer.

However, to efficiently implement such sharing of fog compute capacity, traditionally independent fog nodes must be federated together via coordinated orchestration and control of the distributed fog nodes over the network infrastructure. Such a group of linked fog nodes with non-application specific resources is called a federated fog network [26]. The reach of such federation of fog nodes should be limited (i.e. within a metropolis, suburb, town, or city) to ensure the low latency goal of the fog computing paradigm is not compromised. This is because the expected and emerging fog applications will exhibit a range of delay sensitivities. Certain mission critical fog applications such as vehicle-to-everything and industrial process control may require sub-milliseconds (end-to-end) delay, and at most a delay of 20ms or less to be practical while others may be moderately sensitive to delay [26].

The rising demand for the federation of distributed fog nodes (fog network) is motivating the emergence of service-oriented access models for computing resource in the fog computing era in the form of Fog Infrastructure as a Service (FIaaS) and Fog as a Service (FaaS) [26], [27]. FIaaS provides infrastructure on-demand as in traditional cloud IaaS service offering while FaaS provides platforms or software that support fog apps as in traditional cloud server-less (PaaS or SaaS) service offerings. The adoption of service-oriented consumption of federated fog computing capacity can streamline the number of fog nodes required. Fog networks can also enable new revenue sources which can offset the TCO incurred by providers that deploy fog computing nodes [26]. Furthermore, the federation of distributed fog nodes should support scale-out (horizontal scaling) via on-demand integration and deployment of new fog nodes to support applications latency and computation requirements. The ecosystem of federated distributed fog nodes can also adopt the emerging concept of server resource disaggregation for better efficiency as the concept matures. This paper studies the impact of using disaggregated server architectures in a metro-access network of distributed fog nodes.

### *B. Resource Disaggregation*

Resource disaggregation proposes the separation of the resource components of traditional computing servers into physical or logical pools of homogenous resource types i.e. homogenous resources are in the same pool. The concept addresses the problems of resource stranding and fragmentation which inhibit efficiencies, modularity, agility, and effective resource lifecycle management in traditional computing infrastructure. Resource disaggregation solves these problems by relaxing the rigid physical resource utilization boundary of traditional servers to permit on-demand use of un-collocated resource components (via networks) to form logical computing hosts. Disaggregation of the server intrinsic resource components can be performed at different scales i.e. rack-scale, pod-scale and DC-scale [16], [28], [29]. Furthermore, the concept of resource disaggregation can also be implemented physically or logically over a corresponding physical hardware. On the one hand, physical resource disaggregation ensures that the allocation of physical resource components leads to the creation of physical pools (nodes/racks/pods) of homogenous resource components. On the other hand, logical resource disaggregation virtually implements the concept of resource disaggregation on homogenous and/or heterogeneous resource pools to support on-demand creation of logical servers to run applications.

Resource disaggregation promises significant improvements in the efficiency of computing infrastructure at both the fog and cloud computing tiers of the cloud of things continuum. However, such infrastructure requires software defined (SD) techniques and appropriate network connectivity to maximize the potential benefits enabled by resource disaggregation. In recent times, the adoption of SD-techniques in the computing environment has been popularized via the adoption of concepts such as SDDC, SD-infrastructure, SD-computing, SD-storage and SD-network in both small-scale and large-scale computing infrastructure of enterprises and hyper-scale cloud DC providers respectively [30]. A computing infrastructure that implements resource disaggregation builds on the efficiencies, flexibility and agility enabled by these SD-techniques to enable even greater efficiency in a fully composable computing infrastructure. An appropriate network topology that supports on-demand creation of logical hosts from disaggregated underlying resource components is essential for practical implementation of any form of resource disaggregation [31]. This is because resource disaggregation exposes high bandwidth and low latency intra-host communication of traditional servers into the DC network which even now carries high volumes of dynamic north-south and east-west traffic in DCs.

### *C. Related Works*

Comprehensive studies of energy efficient communication networks have been done in [32]–[37] as a result of increasing adoption of digital solutions and services, which are often domiciled in remote hyper-scale DCs of cloud computing services providers. However, the recent introduction of the fog computing layer motivates an extension of similar studies to the fog computing layer. The authors of [18]–[20] have also conducted extensive studies on how the introduction of the fog computing layer can enable significant power savings in the cloud of things architecture relative to the adoption of a 2 tier architecture. The work in [9] also showed that a 3 tier cloud of things architecture is better than a 2 tier cloud of things architecture especially when energy consumption and latency are used as comparison metrics; hence, a 3-tier cloud of things architecture is adopted in this paper. However, the goal is to minimize the number of computational nodes in the lowest tier of the 3-tier architecture by employing the concept of server disaggregation for the first time in this context.

In [7], the authors gave a description of the fog computing architecture while highlighting the benefits and limitations of using the fog layer as a middleware between the cloud and numerous IoT devices. Performance evaluation showed that the execution of large tasks experiences significant processing delay in the fog computing layer. Hence, scaling fog computing capacity at the expense of higher financial cost is required. In [26], a platform which orchestrates distributed fog nodes over the network to form fog networks that support on-demand deployment of applications and services was proposed. The authors in [38] introduced a programming model for present and emerging geo-distributed, massive and latency-sensitive applications. The PaaS programming model called "Mobile Fog" provides a simplified programming abstraction and supports on-demand scaling of applications at runtime to use resources in cloud or fog computing tiers. The programming model was validated using two latency sensitive applications. The results showed that the introduction of a fog computing tier closer to application users enabled significant reductions in network latency and core network traffic when applications query ranges are small. The authors also encouraged dynamic scaling of fog application instance to effectively handle highly skewed application.

In [10], the authors proposed a high-level policy for placing applications in the fog computing era based on application latency requirements only. The policy is generic and does not consider the impact of factors such as energy efficiency, resource utilization, networks and disaggregation on optimal application placement. Workload offloading and workload assignment are two different approaches used to study the minimization of response time in fog computing era [11]. Given a set of fog workloads, the workload assignment approach attempts to assign such workloads to fog computing nodes while optimizing a specific cost such as response time or energy. On the other hand, the workload offloading approach aims to design policies that offload fog computing requests to other fog nodes or to the cloud while optimizing a specific cost. In [11], [39], the authors explored the workload offloading approach by proposing a general delay minimizing policy for fog nodes to offload IoT application requests to other fog nodes or to forward these requests to the cloud. Using an analytical model, the authors evaluated the policy and showed that the proposed policy reduces response time for IoT applications. The treatment in [23] adopted the workload assignment approach by using a mathematical model formulation to compare fog computing and traditional cloud computing in the IoT era using criteria such as power consumption, cost and latency. Results from the model showed that a cloud of things architecture with fog computing as middleware outperforms an architecture without fog computing layer only when there are many latency-sensitive applications. Otherwise, it is better not to deploy the fog computing layer to ensure that overhead cost, with little or no performance benefit, is not incurred. In [40], the authors studied the trade-off between power consumption and latency in the fog-cloud computing system by formulating a workload placement problem i.e. using the workload assignment approach. Results obtained from the formulated problem via simulation showed that fog computing can significantly improve the performance of cloud computing by reducing communication latency.

The workload assignment approach is adopted in this paper by formulating a MILP model to assign varying classes of interactive fog applications. This is because the workload offloading approach may be less appropriate for interactive workloads with stringent delay requirements. A heuristic is proposed from the insights obtained from the MILP model. In contrast to existing literatures, this work is focused on the fog computing tier only and its associated access and metro networks. The things and cloud computing layers of the cloud of thing architecture are not explicitly considered. However, the impact of cloud destined network traffic on the overall performance of the fog computing system is modelled and a cost is associated with fog servers that are deployed very close to the things layer. Like the works of the authors in [18]–[20], [23], this work explores power consumption, cost and latency in the fog computing layer. However, the novelty of this work is that it also considers disaggregation of fog servers and considers different classes of delay sensitive (interactive) applications. The gains that can be achieved over traditional practice if fog computing nodes adopt DSs over TSs are evaluated. Other factors that may influence performance in such a setup are also studied.

In contrast to [15] where the authors focused on the design of an energy efficient network for DCs with disaggregated servers, we do not consider DC networks in this paper to simplify our evaluation scenario. Moreover, in a previous work [31], we formulated a MILP model that placed workloads energy efficiently in composable DC networks and compared the performance of disaggregation at rack-scale and pod-scale over selected electrical, hybrid and optical DC networks. The results showed that the optical network enabled optimal energy efficiency. It was also reported that logical resource disaggregation at rack-scale is sufficient to deliver similar efficiencies as physical disaggregation at pod-scale while also offering greater flexibility. Hence, this paper exits the domain of DC networks to focus on the application of the resource disaggregation concept in the fog computing layer of the cloud of things architecture. Furthermore, we adopt logical disaggregation of TSs as a representation of the resource disaggregation concept.

## III. MILP Model for Fog Applications Placement

### A. System Setup

The placement of fog computing applications in traditional and disaggregated fog computing nodes is explored given constraints such as resource availability, application performance, energy efficiency and resource utilization. Although fog applications may require functions across the different tiers of the cloud of things architecture, only functions that can be performed in the fog computing tier, such as data pre-processing, aggregation, and intelligence (e.g. facial recognition), are considered. While the primary aim of fog computing concept is to host application and services on the nearest fog devices, this work assumes that some fog computing sites are better than others based on criteria (such as energy efficiency) if the application's delay requirement is satisfied. The primary assumption is that it is better to host fog applications in central locations at the network edge to improve overall energy efficiency. Otherwise, local fog node must be provisioned for such applications. Hence, it is assumed that preferred fog computing nodes in the fog infrastructure are either specialized wired/wireless network equipment which can support generic application, or an existing computing infrastructure owned by an enterprise or a network provider as shown in Fig. 1. Such preferred fog computing sites may also support mission critical traditional applications i.e. virtual machine (VM) and/or virtual network functions (VNF) required by their owners. Therefore, each traditional app (TA) is associated with specific fog computing sites in the network topology and the spare computing resource capacity in such sites, which is not used to support VM/VNF of TA, is made available to the pool of federated fog computing capacity. The use of both traditional and disaggregated computing infrastructure in such fog sites, which have been integrated into a fog network, are compared.

Furthermore, a scenario where the fog computing layer must process all delay sensitive application is considered, this is because such applications cannot be supported by the centralized cloud computing architecture. If a delay sensitive application is not provisioned in the fog network, a local fog node must be provisioned at the source of the request for that application. Every provisioned instance of any fog application leads to corresponding pre-processing and post-processing traffic in the network. Regular network traffic from (to) each access node comprises of traditional network traffic and the traffic from (to) other applications which are processed in centralized cloud computing node. We only consider delay in metropolitan area network (MAN) and access network. Computation (processing) delay is not considered at each fog node. We consider a scenario where the cumulative computation requirements of all users served by an instance of a given fog application is less than the computing capacity provisioned for that fog application by default. Hence, minimal computation delay is suffered. Delay estimation considers only link communication delay which is a sum of propagation and congestion delay experienced on each link in the network topology.

A scenario where fog applications traffic between a fog computing site and end-users follows a single path is considered to simplify delay calculations. Such single low latency path is provisioned by the network service provider to support interactive fog apps created in the fog network. Similar to the work of the authors in [40], a maximum delay threshold is adopted for interactive fog apps during delay-aware placement. As there are different use cases at the access network layer for a federation of fog nodes in the network, it is expected that the network components traversed by network traffic in the access layer will also differ according to each use case. Fig. 1 also gives illustrations of such use cases and their corresponding access network architecture.

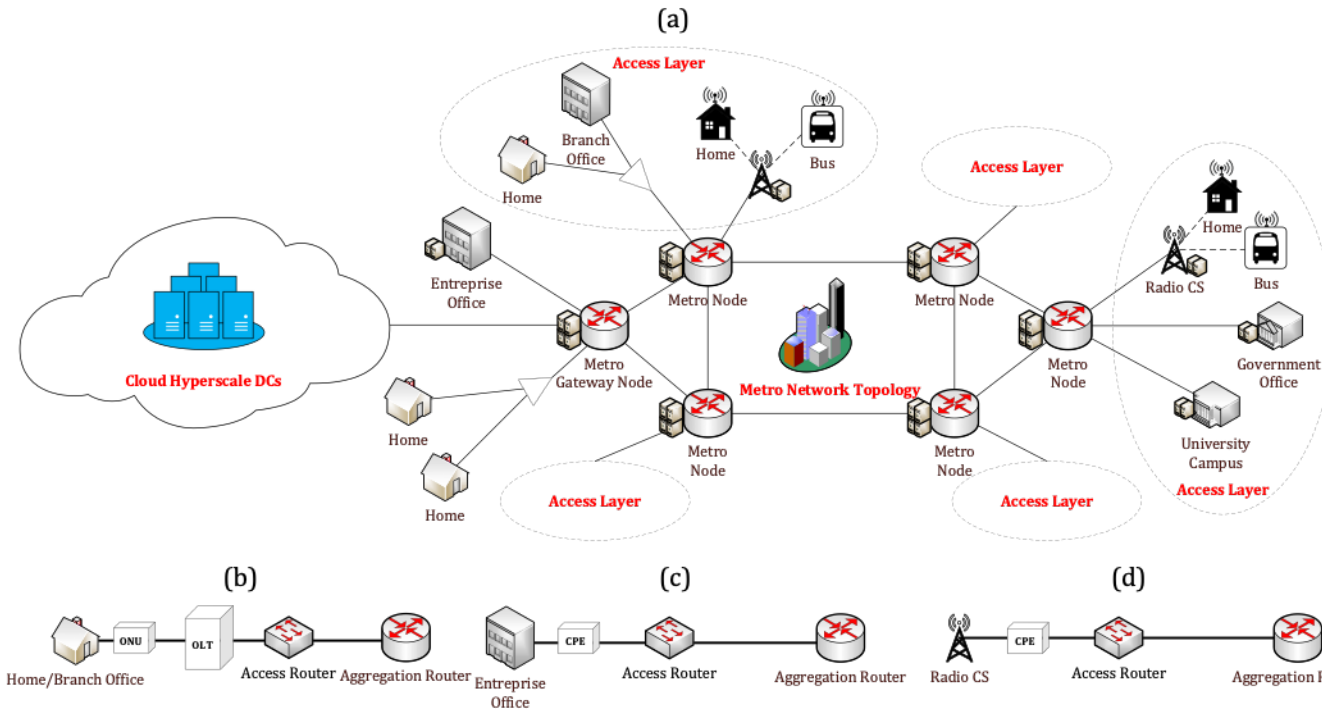

Fig. 1. (a) Metro fog network. (b) Residential or branch office use case. (c) Enterprise use case. (d) Mobile back-haul use case.

### B. MILP Model Description

In this sub-section, a MILP model that efficiently assigns instances of interactive applications in distributed fog computing nodes within a MAN topology is presented. The model minimizes network power consumption, fog computing power consumption, resulting power consumption of rejected fog applications and the approximated total queuing delay incurred in the network. Given: (i) a MAN topology comprising of sets of metro and access network nodes and corresponding inter-connecting physical link capacities as illustrated in Fig. 1; (ii) the availability of fog computing capacity in selected fog sites/network nodes in the MAN topology; and (iii) the locations of clusters of end-users/IoT-devices with explicit demand for an instance of fog applications; the model determines the number of instances of each fog app that can be provisioned and the optimal location of each provisioned instance while enforcing defined constraints. The data traffic at a given node is proportional to the number of users in that node. Hence, both pre-processing and post processing data traffic of each fog app instance are defined in Gbps per user. Furthermore, only an instance of a given fog app is allocated to all users of that fog app in each access node. The model parameters and variables are given as follows and linear approximations are made as required to ensure linearity.

**Network sets and parameters:**

| | |
|---|---|
| $N$ | Set of network nodes |
| $NB_m$ | Set of neighbor nodes of network node $m \in N$ |
| $MN$ | Set of metro network nodes, $MN \subseteq N$ |
| $MNB_m$ | Set of neighbor metro nodes of metro node $m \in MN$ |
| $GW$ | Set of gateway metro nodes in metro network topology, $GW \subseteq MN$ |
| $AN$ | Set of access network nodes $AN \subseteq N$ |
| $ANB_a$ | Set of neighbor metro nodes of access network node $a \in AN$ |
| $ANCPE_a$ | $ANCPE_a = 1$ if access network node $a \in AN$ has a consumer premises equipment. Otherwise, $ANCPE_a = 0$. |
| $ANPON_a$ | $ANPON_a = 1$ if access network node $a \in AN$ has a PON ONU. Otherwise, $ANPON_a = 0$. |
| $B_{mn}$ | Bandwidth of physical link $(m,n)$ $m \in N, n \in N_m$ |
| $\Delta_{an}$ | $\Delta_{an} = 1$ if node $n \in N$ is an access network node $a \in AN$. Otherwise, $\Delta_{an} = 0$ |
| $MNT_{mn}$ | Regular traffic on physical link $(m,n)$ $m \in N, n \in NB_m$ |
| $PD_{mn}$ | Propagation delay on physical link $(m,n)$ $m \in N, n \in NB_m$ |
| $LP_{mn}$ | Set of linear pieces (linear approximations) used to linearize the delay curve of the delay experienced on link $(m,n)$ $m \in N, n \in NB_m$ |
| $Rate_{mnq}$ | Rate of linear piece $q \in LP_{mn}$ of the linear approximation of the delay experienced on link $(m,n)$ $m \in N, n \in NB_m$ |
| $I_{mnq}$ | Intercept of linear piece $q \in LP_{mn}$ of the linear approximation of the delay experienced on link $(m,n)$ $m \in N, n \in NB_m$ |
| $LU_{mn}$ | Upper bound of queuing delay experienced on link $(m,n)$ $m \in N, n \in NB_m$ |
| $CPEPC$ | Consumer premises equipment power consumption |
| $MANepb$ | Metro Ethernet access switch energy per bit |
| $MAGNepb$ | Metro Ethernet aggregation switch energy per bit |
| $ONUPC$ | PON ONU power consumption |
| $OLTepb$ | PON OLT energy per bit |
| $\delta$ | Queuing penalty of the network topology in Watt per second. |

**Fog applications sets and parameters:**

| | |
|---|---|
| $F$ | Set of fog apps |
| $T$ | Set of traditional fog apps, $T \subseteq F$ |
| $E$ | Set of emerging fog apps, $E \subseteq F$ |
| $FCD_f$ | Compute resource demand of fog app $f \in F$ |
| $FMD_f$ | Memory resource demand of fog app $f \in F$ |
| $FSD_f$ | Storage resource demand of fog app $f \in F$ |
| $FUP_e$ | Uplink data rate per user of emerging fog app $e \in E$ |
| $FDW_e$ | Downlink data rate per user of emerging fog app $e \in E$ |
| $FSRC_{tn}$ | $FSRC_{tn} = 1$ if traditional fog app $t \in T$ is associated with network node $n \in N$. Otherwise, $FSRC_{tn} = 0$ |
| $FDST_{en}$ | $FDST_{en} = 1$ if node $n \in GW$ is the gateway node of emerging fog app $e \in E$. Otherwise, $FDST_{en} = 0$ |
| $FinAN_{ea}$ | $FinAN_{ea} = 1$ if users in node $a \in AN$ request an instance of emerging fog app $e \in E$. Otherwise, $FinAN_{ea} = 0$. |
| $FUinAN_{ea}$ | Number of users in node $a \in AN$ requesting an instance of emerging fog app $e \in E$. |
| $ED_e$ | Emerging fog app $e \in E$ maximum delay threshold. |
| $\gamma$ | Cost coefficient of power consumed because of traditional fog apps rejection. |
| $\emptyset$ | Cost coefficient of power consumed because of emerging fog apps rejection. |

**Fog computing nodes sets and parameters:**

| | |
|---|---|
| $C$ | Set of CPU resource components. |
| $M$ | Set of memory resource components. |
| $S$ | Set of storage resource components. |
| $CPU_c$ | Capacity of CPU component $c \in C$ |
| IC | Idle power consumption as a fraction of the maximum CPU power consumption. |
| $CPmax_c$ | Maximum power consumption of CPU $c \in C$ |
| $\Delta C_c$ | Power factor of CPU module $c \in C$; $\Delta C_c = \frac{CPmax_c - IC\ CPmax_c}{CPU_c}$ |
| $RAM_m$ | Capacity of memory component $m \in M$ |
| IM | Idle power consumption as a fraction of the maximum memory power consumption. |
| $MPmax_m$ | Maximum power consumption of memory component $m \in M$ |
| $\Delta M_m$ | Power factor of memory component $m \in M$; $\Delta M_m = \frac{MPmax_m - IM\ MPmax_m}{RAM_m}$ |
| $STOR_s$ | Capacity of storage component s $\in S$ |
| IS | Idle power consumption as a fraction of the maximum storage power consumption. |
| $SPmax_s$ | Maximum power consumption of storage component $s \in S$ |
| $\Delta S_s$ | Power factor of storage component $s \in S$; $\Delta S_s = \frac{SPmax_s - IS\ SPmax_s}{STOR_s}$ |
| $CPmax$ | Peak power consumption of the CPU component with highest power consumption. |
| $\Delta Cmax$ | Power factor of the CPU component with highest power consumption. |
| $MPmax$ | Peak power consumption of the memory component with highest power consumption. |
| $\Delta Mmax$ | Power factor of the memory component with highest power consumption. |
| $SPmax$ | Peak power consumption of the storage component with highest power consumption. |
| $\Delta Smax$ | Power factor of the storage component with highest power consumption. |
| $CN$ | Set of computing nodes |
| $CinCN_{cx}$ | $CinCN_{cx} = 1$ if CPU component $c \in C$ is placed at computing node $x \in CN$. Otherwise, $CinCN_{cx} = 0$ . |
| $MinCN_{mx}$ | $MinCN_{mx} = 1$ if memory component $m \in M$ is placed at computing node $x \in CN$. Otherwise, $MinCN_{mx} = 0$. |
| $SinCN_{sx}$ | $SinCN_{sx} = 1$ if storage component $s \in S$ is placed at compute node $x \in CN$. Otherwise, $SinCN_{sx} = 0$. |
| $CNinNode_{xn}$ | $CNinNode_{xn} = 1$ if compute node $x \in CN$ is placed at network node $n \in N$. Otherwise, $CNinNode_{xn} = 0$. |
| $CinNode_{cn}$ | $CinNode_{cn} = 1$ if CPU component $c \in C$ is placed at network node $n \in N$. Otherwise, $CinNode_{xn} = 0$. |
| $\mu$ | A large enough number. |

**Variables:**

| | |
|---|---|
| $FCL_{fc}$ | $FCL_{fc} = 1$ if an instance of fog app $f \in F$ is in CPU component $c \in C$. Otherwise, $FCL_{fc} = 0$. |
| $FML_{fm}$ | $FML_{fm} = 1$ if an instance of fog app $f \in F$ is in memory component $m \in M$. Otherwise, $FML_{fm} = 0$. |
| $FSL_{fs}$ | $FSL_{fs} = 1$ if an instance of fog app $f \in F$ is in storage component $s \in S$.Otherwise, $FSL_{fs} = 0$. |
| $CA_c$ | $CA_c = 1$ if CPU component $c \in C$ is active. Otherwise, $CA_c = 0$. |
| $MA_m$ | $MA_m = 1$ if memory component $m \in M$ is active. Otherwise, $MA_m = 0$. |
| $SA_s$ | $SA_s = 1$ if storage component $s \in S$ is active. Otherwise, $SA_s = 0$. |
| $X_{eca}$ | $X_{eca} = 1$ if the instance of emerging fog app $e \in E$ in CPU component $c \in C$ is allocated to users in access node $a \in AN$. Otherwise, $X_{eca} = 0$. |
| $V_{ea}$ | Number of instances of emerging fog app $e \in E$ allocated to users of that fog app in access node $a \in AN$. |
| $\varphi_{esa}$ | $\varphi_{esa} \geq 1$ if an instance of emerging fog app $e \in E$ in node $s \in N$ is allocated to users of that app in access node $a \in AN$. Otherwise, $\varphi_{esa} = 0$. |
| $\Psi_{sd}$ | Post-processing traffic from instances of all fog apps in network node $s \in N$ to gateway node $d \in GW$. |
| $\Omega_{sdec}$ | Pre-processing traffic from users in node $s \in N$ to node $d \in N$ that hosts an instance of emerging fog app $e \in E$ placed in CPU component $c \in C$ in node $d \in N$. |
| $\Phi_{sdec}$ | Post-processing traffic from compute node $s \in N$ to users in node $d \in N$. Node $s \in N$ hosts an instance of emerging fog app $e \in E$ placed in CPU component $c \in C$. The instance of emerging fog app $e \in E$ placed in CPU component $c \in C$ was allocated to users in node $d \in N$. |
| $L_{sdec}$ | Traffic from node $s \in N$ to node $d \in N$ due to the presence of emerging fog app $e \in E$ in CPU component $c \in C$. |
| $H_{mn}^{sd}$ | Volume of $\Psi_{sd}$ traffic routed on physical link $(m, n)$ |
| $\lambda_{mn}$ | Volume of cloud bound traffic on physical link $(m, n)$. |
| $M_{mn}^{sdec}$ | Flow of latency sensitive traffic (emerging fog applications traffic) $L_{sdec}$ on physical link $(m, n)$ |
| $MB_{mn}^{sdec}$ | $MB_{mn}^{sdec} = 1$ if a flow of $L_{sdec}$ is present on physical link $(m, n)$. Otherwise $MB_{mn}^{sdec} = 0$. |
| $\Lambda_{mn}$ | Volume of latency sensitive traffic on physical link $(m, n)$ |
| $\Gamma_{mn}$ | Total traffic on physical link $(m, n)$. |
| $STA_t$ | State of traditional fog app $t \in T$. |
| $TCRTA$ | Total cost of rejected traditional fog apps in Watts. |
| $TCREA$ | Total cost of rejected emerging fog app in Watts. |
| $\alpha_t$ | Power penalty because of rejecting traditional fog app $t \in T$ in Watts. |
| $\beta_e$ | Power penalty because of rejecting emerging fog app $e \in E$ in Watts. |
| $W_{mn}$ | M/M/1 queuing delay experienced on physical link $(m, n)$ |
| $TL_{mn}^{sdec}$ | $TL_{mn}^{sdec}$ is the queuing delay experienced by flow $L_{sdec}$ on physical link $(m, n)$ on the path selected for the flow. |
| $PD_{mn}^{sdec}$ | $PD_{mn}^{sdec}$ is the propagation delay experienced by flow $L_{sdec}$ on physical link $(m, n)$ on the path selected for the flow. |
| $WL_{sdec}$ | Total delay of flow $L_{sdec}$ on all physical links. Sum of congestion experienced on physical links (queuing delay) and propagation delay on physical links on the path. |
| $RTD_{sdec}$ | Round trip delay between a node containing users of an emerging fog app and the network node hosting the instance assigned to the users. |
| $TDL$ | Approximated total queuing delay experienced on physical links of the network topology. |

Delay related variables in the MILP model are derived as follows:

$$PD_{mn}^{sdec} = PD_{mn} MB_{mn}^{sdec} \quad (1)$$

$$\forall s, d \in N, \forall e \in E, \forall c \in C, \forall m \in N, \forall n \in NB_m$$

Equation (1) gives the propagation delay experienced by flow $L_{sdec}$ on physical link $(m, n)$ on the path selected for the flow.

$$WL_{sdec} = \sum_{m \in N} \sum_{n \in N_m} (TL_{mn}^{sdec} + PD_{mn}^{sdec}) \quad (2)$$

$$\forall s, d \in N, \forall e \in E, \forall c \in C$$

Equation (2) gives the total delay experienced by flow $L_{sdec}$ on all physical links, it is a sum of congestion delay and propagation delay experienced on all physical links on the path.

$$RTD_{sdec} = WL_{sdec} + WL_{dsec} \quad (3)$$

$$\forall s, d \in N, \forall e \in E, \forall c \in C$$

Equation (3) gives the round-trip delay experienced between a node with users of an emerging fog app and the network node hosting the instance assigned to the users at that node.

$$TDL = \sum_{m \in N} \sum_{n \in NB_m} W_{mn} \quad (4)$$

Equation (4) gives the approximated total delay experienced on all physical links of the network topology.

The following equations present the derivation of variables that aid the calculation of network traffic and power consumption.

$$\varphi_{esa} = \sum_{c \in C} X_{eca}\, CinNode_{cs} \quad (5)$$

$$\forall s \in N, \forall a \in AN, \forall e \in E$$

Equation (5) gives the variable $\varphi_{esa}$ which depends on the placement of emerging fog application. A scenario where the compute capacity of an instance of emerging fog app $e \in E$ is greater than the maximum compute capacity required by the cluster of users of that app in all access node is considered. Hence, $\varphi_{esa} > 1$ is avoided.

$$\Omega_{sdec} = \sum_{a \in AN} X_{eca} FUP_e FUinAN_{ea}\, \Delta_{as}\, CinNode_{cd} \quad (6)$$

$$\forall s, d \in N, \forall e \in E, \forall c \in C$$

Equations (6) gives the pre-processing traffic between users in access nodes and instances of emerging fog applications placed in fog computing sites.

$$\Phi_{sdec} = \sum_{a \in AN} X_{eca}\, FDW_e\, FUinAN_{ea}\, \Delta_{ad}\, CinNode_{cs} \quad (7)$$

$$\forall s, d \in N, \forall e \in E, \forall c \in C$$

Equations (7) gives the post-processing traffic between users in access nodes and instances of emerging fog applications placed in fog computing sites.

$$L_{sdec} = \Omega_{sdec} + \Phi_{sdec} \quad (8)$$

$$\forall s, d \in N, \forall e \in E, \forall c \in C$$

Equation (8) gives the total traffic between a pair of nodes due to the creation of an instance of an emerging fog app in the fog network.

$$\Psi_{sd} = \sum_{a \in AN} \sum_{e \in E} \varphi_{esa}\, FDW_e\, FUinAN_{ea}\, FDST_{ed} \quad (9)$$

$$\forall s \in N, \forall d \in GW$$

Equation (9) gives the post-processing traffic between instances emerging fog applications placed in fog computing sites and gateway metro nodes.

$$\Lambda_{mn} = \sum_{s \in N} \sum_{d \in N} \sum_{e \in E} \sum_{c \in C} M_{mn}^{sdec} \quad (10)$$

$$\forall m \in N, \forall n \in NB_m$$

Equation (10) gives the delay sensitive traffic routed over a physical link by summing all the delay sensitive flows over the link.

$$\lambda_{mn} = \sum_{s \in N} \sum_{d \in N} H_{mn}^{sd} + MNT_{mn} \quad (11)$$

$$\forall m \in N, \forall n \in NB_m$$

Equation (11) gives the delay tolerant traffic routed over a physical link by summing all the delay tolerant flows over the link with the given regular traffic on that physical link.

$$\Gamma_{mn} = \Lambda_{mn} + \lambda_{mn} \quad (12)$$

$$\forall m \in N, \forall n \in NB_m$$

Equation (12) gives the total traffic over a link by summing delay sensitive and delay tolerant traffic routed over the link.

Total network power consumption $TNetPC$ is given by:

$$TNPC = TMNETPC + TANETPC \quad (13)$$

$TMNETPC$ is the total metro network power consumption

and is given by

$$TMNETPC = \sum_{m \in MN} (A_m + B_m + C_m)\, MAGNepb \quad (14)$$

where $A_m$ is the traffic relayed by a metro node $m$ and is given by

$$A_m = \sum_{s \in N} \sum_{d \in N} \sum_{n \in NB_m} H_{mn}^{sd} + \sum_{n \in NB_m} \sum_{s \in N} \sum_{d \in N} \sum_{e \in E} \sum_{c \in C} M_{mn}^{sdec} \quad (15)$$

$$\forall m \in MN, s, d \in N, s \neq m, d \neq m$$

$B_m$ is the traffic received by a metro node $m$ and is given by

$$B_m = \sum_{s \in N} \sum_{n \in NB_m} H_{nm}^{sm} + \sum_{s \in N} \sum_{n \in NB_m} \sum_{e \in E} \sum_{c \in C} M_{nm}^{smec} + \sum_{n \in NB_m} MNT_{nm} \quad (16)$$

$$\forall m \in MN$$

and finally $C_m$ is the traffic transmitted by a metro node $m$ and is given by

$$C_m = \sum_{d \in GW} \sum_{n \in NB_m} H_{mn}^{md} + \sum_{d \in N} \sum_{n \in NB_m} \sum_{e \in E} \sum_{c \in C} M_{mn}^{mdec} + \sum_{n \in NB_m} MNT_{mn} \quad (17)$$

$$\forall m \in MN$$

$TANETPC$ is the total access network power consumption and is given by

$$TANETPC = \sum_{a \in AN} (CPEPC\ ANCPE_a + ONUPC\ ANPON_a) + \sum_{a \in AN} \sum_{m \in ANB_a} (\Gamma_{am} + \Gamma_{ma})(MANepb + ANPON_a\ OLTepb) \quad (18)$$

The total fog computing power consumption ($TFPC$) is given by:

$$TFPC = TCPC + TMPC + TSPC \quad (19)$$

where $TCPC$ is the total power consumption of CPU resources in fog network and is given by

$$TCPC = \sum_{c \in C} \left( IC\ CPmax_c\ CA_c + \sum_{f \in F} \Delta C_c\ FCL_{fc}\ FCD_f \right) \quad (20)$$

$TMPC$ is the total power consumption of memory resources in fog network and is given by

$$TMPC = \sum_{m \in M} \left( IM\ MPmax_m\ MA_m + \sum_{f \in F} \Delta M_m\ FML_{fm}\ FMD_f \right) \quad (21)$$

and $TSPC$ is the total power consumption of storage resources in fog network and is given by

$$TSPC = \sum_{s \in S} \left( IS\ SPmax_s\ SA_s + \sum_{f \in F} \Delta S_s\ FSL_{fs}\ FSD_f \right) \quad (22)$$

The total cost of rejected traditional fog apps in the distributed fog network is given by

$$TCRTA = \sum_{t \in T} \alpha_t \quad (23)$$

where

$$\alpha_t = \big((IC\ CPmax + \Delta Cmax\ FCD_t) + (IM\ Pmax + \Delta Mmax\ FMD_t) + (IS\ SPmax + \Delta Smax\ FSD_t)\big) STA_t \quad (24)$$

$$\forall t \in T$$

where the state of traditional application t is given by

$$STA_t = \sum_{c \in C} (1 - FCL_{tc}) \quad (25)$$

$$\forall t \in T$$

The total cost of rejected emerging fog apps in the distributed fog network is given by

$$TCREA = \sum_{e \in E} \beta_e \quad (26)$$

where

$$\beta_e = \big((IC\ CPmax + \Delta Cmax\ FCD_e) + (IM\ MPmax + \Delta Mmax\ FMD_e) + (IS\ SPmax + \Delta Smax\ FSD_e)\big) \sum_{a \in AN} (1 - V_{ea}) \quad (27)$$

$$\forall e \in E$$

where $V_{ea}$ indicates if the emerging application $e$ requested by node $a$ has been provisioned or not, and is given by

$$V_{ea} = \sum_{c \in C} X_{eca} \quad (28)$$

$$\forall e \in E, \forall a \in AN$$

It is important to note that the cost of rejecting an app is defined as the maximum power consumed if it is accepted, i.e. the power consumption in the case where inactive components must be turned-on to support the app.

The model is defined as follows:

**Objective**: Minimize

$$TNPC + TFPC + \gamma\ TCRTA + \emptyset\ TCREA + \delta\ TDL \quad (29)$$

The objective of the model is to minimize a weighted sum of the TNPC, TFPC, the total cost of rejected traditional, emerging fog applications and the cost of approximated total delay in the network as given by the expression in (29). Note that both traditional and emerging fog apps are provisioned or rejected based on the cost coefficient of terms in the objective. Setting $\gamma$ to a high value ensures that TCRTA is significantly higher than TCREA. Hence, higher priority is given to provisioning traditional fog apps in the objective function. $\emptyset$ can also be varied to increase or decrease the cost of rejected emerging fog apps. The value of $\delta$ dictates the weight of approximated total queuing delay in the objective function. $\delta \ll 1$ represents a network with trivial queuing delay penalty while $\delta \gg 1$ represents a network with significant queuing delay penalty.

**Subject to the following constraints:**

**Fog DC related constraints**

$$\sum_{c \in C} FCL_{fc} = \sum_{m \in M} FML_{fm} \quad \forall f \in F \tag{30}$$

$$\sum_{c \in C} FCL_{fc} = \sum_{s \in S} FSL_{fs} \quad \forall f \in F \tag{31}$$

Constraints (30) and (31) ensure that the number of instances of CPU resources provisioned for any (traditional or emerging) fog app is equal to the number of instances of memory and storage resources provisioned for that app across the distributed fog sites.

$$\sum_{c \in C} FCL_{fc}\, CinCN_{cx} = \sum_{m \in M} FML_{fm}\, MinCN_{mx} \quad \forall f \in F, \forall x \in CN \tag{32}$$

$$\sum_{c \in C} FCL_{fc}\, CinCN_{cx} = \sum_{s \in S} FSL_{fs}\, SinCN_{sx} \quad \forall f \in F, \forall x \in CN \tag{33}$$

Constraints (32) and (33) are the locality constraints when the traditional server architecture is adopted in compute nodes. They ensure that the CPU, memory, and storage components used to provision a given instance of a fog app are in the same compute nodes.

$$\sum_{x \in CN} \sum_{c \in C} FCL_{fc}\, CinCN_{cx}\, CNinNode_{xn} = \sum_{x \in CN} \sum_{m \in M} FML_{fm}\, MinCN_{mx}\, CNinNode_{xn} \quad \forall f \in F, \forall n \in N \tag{34}$$

$$\sum_{x \in CN} \sum_{c \in C} FCL_{fc}\, CinCN_{cx}\, CNinNode_{xn} = \sum_{x \in CN} \sum_{s \in S} FSL_{fs}\, SinCN_{sx}\, CNinNode_{xn} \quad \forall f \in F, \forall n \in N \tag{35}$$

Constraints (34) and (35) are the locality constraints when the disaggregated server architecture is adopted in compute nodes. They ensure that the CPU, memory, and storage components used to provision a given instance of a fog app are in the same fog computing site (network node) but not necessarily in the same compute node.

$$\sum_{x \in CN} \sum_{c \in C} FCL_{tc}\, CinCN_{cx}\, CNinNode_{xn} = FSRC_{tn} \quad \forall t \in T, \forall n \in N \tag{36}$$

Constraint (36) is the workload locality constraint for traditional fog apps associated with a given network node.

$$\sum_{c \in C} FCL_{fc}\, CinCN_{cx} \le 1 \quad \forall f \in F, \forall x \in CN \tag{37}$$

$$\sum_{m \in M} FML_{fm}\, MinCN_{mx} \le 1 \quad \forall f \in F, \forall x \in CN \tag{38}$$

$$\sum_{s \in S} FSL_{fs}\, SinCN_{sx} \le 1 \quad \forall f \in F, \forall x \in CN \tag{39}$$

Constraints (37) - (39) are SLA constraints which ensure robustness of the fog network. They ensure that only an instance of fog app $f$ is provisioned within a given compute node. Hence, the impact of a compute node failure is minimized for a fog app with multiple instances.

$$\sum_{f \in F} FCD_f\, FCL_{fc} \le CPU_c \quad \forall c \in C \tag{40}$$

$$\sum_{f \in F} FMD_f\, FML_{fm} \le RAM_m \quad \forall m \in M \tag{41}$$

$$\sum_{f \in F} FSD_f\, FSL_{fs} \le STOR_s \quad \forall s \in S \tag{42}$$

Constraints (40) - (42) denote resource capacity constraints for each CPU, memory, and storage component in the fog network. They ensure that each resource component's capacity across all fog computing sites is not exceeded.

$$\sum_{f \in F} FCL_{fc} \ge CA_c \quad \forall c \in C \tag{43}$$

$$\sum_{f \in F} FCL_{fc} \le \mu\, CA_c \quad \forall c \in C \tag{44}$$

$$\sum_{f \in F} FML_{fm} \ge MA_m \quad \forall m \in M \tag{45}$$

$$\sum_{f \in F} FML_{fm} \le \mu\, MA_m \quad \forall m \in M \tag{46}$$

$$\sum_{f \in F} FSL_{fs} \ge SA_s \quad \forall s \in S \tag{47}$$

$$\sum_{f \in F} FSL_{fs} \le \mu\, SA_s \quad \forall s \in S \tag{48}$$

Constraints (43) - (48) derive the state of CPU, memory, and storage resources components across all fog computing sites.

**Fog app instance related constraints**

$$\sum_{c\in C} X_{eca} \leq 1 \quad (49)$$

$$\forall\, e \in E, \forall\, a \in AN$$

Constraint (49) ensures that the cluster of users requesting an emerging fog app is assigned at most one instance of that application.

$$\sum_{a\in AN} X_{eca} \geq FCL_{ec} \quad (50)$$

$$\forall\, c \in C, e \in E$$

$$\sum_{a\in AN} X_{eca} \leq \mu\, FCL_{ec} \quad (51)$$

$$\forall\, c \in C, e \in E$$

Constraints (50) and (51) ensure that each instance of an emerging fog app in a CPU component is allocated to one or more user clusters in access nodes. Otherwise, the instance should not be created.

$$X_{eca} = FinAN_{ea}\, V_{ea}\, FCL_{ec} \quad (52)$$

$$\forall\, c \in C, e \in E, a \in AN$$

$$X_{eca} \leq FinAN_{ea}\, FCL_{ec} \quad (53)$$

$$\forall\, c \in C, e \in E, a \in AN$$

$$X_{eca} \leq FinAN_{ea}\, V_{ea} \quad (54)$$

$$\forall\, c \in C, e \in E, a \in AN$$

$$X_{eca} \geq FinAN_{ea}(V_{ea} + FCL_{ec}) - 1 \quad (55)$$

$$\forall\, c \in C, e \in E, a \in AN$$

Constraint (52) derives $X_{eca}$ which gives the instance of an emerging fog app in a CPU that is assigned to users of that fog app in an access node. $X_{eca} = 1$ if and only if users of an emerging fop app are present an access node, an instance of that fog app is in a CPU component and that instance has been assigned to users of that fog app in the corresponding access node. Constraints (53) - (55) implement linearly the product of parameter and variables given in Constraint (52).

**Network related constraints**

$$\sum_{n\in NB_m} H_{mn}^{sd} - \sum_{n\in NB_m} H_{nm}^{sd} = \begin{cases} \Psi_{sd} & m = s \\ -\Psi_{sd} & m = d \\ 0 & otherwise \end{cases} \quad (56)$$

$$\forall\, s, m \in N, \forall\, d \in GW : s \neq d$$

Constraint (56) enforces flow conservation for post-processing traffic to the cloud in the physical layer of the network.

$$\sum_{n\in NB_m} M_{mn}^{sdec} - \sum_{n\in NB_m} M_{nm}^{sdec} \quad (57)$$

$$= \begin{cases} L_{sdec} & m = s \\ -L_{sdec} & m = d \\ 0 & otherwise \end{cases}$$

$$\forall\, s, d, m \in N, e \in E, c \in C : s \neq d$$

Constraint (57) enforces flow conservation for delay sensitive flows in the physical layer of the network.

$$M_{mn}^{sdec} \geq MB_{mn}^{sdec} \quad (58)$$

$$\forall\, s, d, m, n \in N, e \in E, c \in C : s \neq d$$

$$M_{mn}^{sdec} \leq \mu\, MB_{mn}^{sdec} \quad (59)$$

$$\forall\, s, d, m, n \in N, e \in E, c \in C : s \neq d$$

Constraints (58) and (59) give the binary equivalent of $M_{mn}^{sdec}$.

$$\sum_{n\in NB_m} MB_{mn}^{sdec} \leq 1 \quad (60)$$

$$\forall\, s, d, m \in N, e \in E, c \in C : s \neq$$

Constraint (60) ensures that the flow $L_{sdec}$ is not bifurcated over multiple paths.

$$\Gamma_{mn} \leq B_{mn} \quad (61)$$

$$\forall\, m \in N, n \in NB_m$$

Constraint (61) enforces capacity constraint on each physical link $(m, n)$.

**Network delay related constraints**

$$TL_{mn}^{sdec} = W_{mn}\, MB_{mn}^{sdec} \quad (62)$$

$$\forall\, s, d, m \in N, n \in NB_m, e \in E, c \in C : s \neq d$$

$$TL_{mn}^{sdec} \leq LU_{mn}\, MB_{mn}^{sdec} \quad (63)$$

$$\forall\, s, d, m \in N, n \in NB_m, e \in E, c \in C : s \neq d$$

$$TL_{mn}^{sdec} \leq W_{mn} \quad (64)$$

$$\forall\, s, d, m \in N, n \in NB_m, e \in E, c \in C : s \neq d$$

$$TL_{mn}^{sdec} \geq W_{mn} - LU_{mn}\,(1 - MB_{mn}^{sdec}) \quad (65)$$

$$\forall\, s, d, m \in N, n \in NB_m, e \in E, c \in C : s \neq d$$

Constraint (62) estimates the queuing delay experienced by flow $L_{sdec}$ on physical link $(m, n)$ which is given by the product of $W_{mn}$ and $MB_{mn}^{sdec}$. Constraints (63) - (65) linearize Constraint (62). $LU_{mn}$ is the upper bound of the queuing delay experienced on each physical link, it is required to ensure that the delay experienced on a physical link does not exceed a predefined threshold.

$$W_{mn} \geq Rate_{mnq}\, \Gamma_{mn} + I_{mnq} \quad (66)$$

$$\forall\, m \in N, n \in NB_m, q \in LP_{mn}$$

Constraint (66) represent piecewise linear approximation of queuing delay experienced on physical link $(m, n)$. This is required because M/M/1 delay is a non-linear function.

$$RTD_{sdec} \leq ED_e \quad (66)$$

$$\forall\, s, d \in N, e \in E, c \in C : s \neq d$$

Constraint (67) represents the round-trip delay constraint for an emerging fog app $e$. The round-trip delay experienced by an emerging fog app $e$ must not exceed the predefined threshold.

## IV. Evaluation and Results

### A. *Evaluation Scenarios and Input Parameters*

The MILP model described in the previous section is used to study the impact of adopting DS architecture in the fog computing layer of the cloud of things continuum relative to the use of TS architecture. To minimize execution time of the MILP model which grows as the size and complexity of the problem increases, a small network topology comprising of 4 (metro) central offices (COs) and 16 access nodes is adopted as the default evaluation scenario. The access nodes comprise of 4 radio cell sites (CSs), 4 enterprise offices (EOs), which are connected to the metro ring via 40Gbps links, and 8 homes, which are connected to the metro ring via 40Gbps Next-Generation Passive Optical Network 2 (NG-PON2) links. Connected to each metro CO are an EO, a radio CS and two residential houses as illustrated in Fig. 2. By using 40Gbps last mile links between metro and access network nodes, network bottlenecks that can lead to the rejection of emerging fog apps as reported in [17] are avoided.

To further maintain simplicity, the evaluation scenario allocates two servers (or compute nodes comprising of commodity hardware) to each fog computing sites. The fog computing sites comprise of metro COs, EOs, and radio CSs in the network topology as illustrated in Fig. 2. When the TS architecture is adopted, the utilization scope of each server's intrinsic resource components is limited to that server. On the other hand, when the DS architecture is adopted, servers are logically disaggregated to expand the utilization scope of the intrinsic resources present in each server at fog computing sites. However, access to such disaggregated resource components is limited to the corresponding fog computing site hosting each component. A common configuration is adopted for all servers distributed across the metro network topology, each server comprises of one CPU, one memory and one storage resource components. The characteristics of each compute resource component used to evaluate the MILP model are given in Table I. The power consumption profile of each compute resource component comprises of an idle portion and another portion that is linearly dependent on the component's utilization. By default, servers are not allocated to residential houses because reduction of the number of computing resources deployed at the extreme edge of the network is desired. Moreover, it is unusual to have large computing capacity in residential houses.

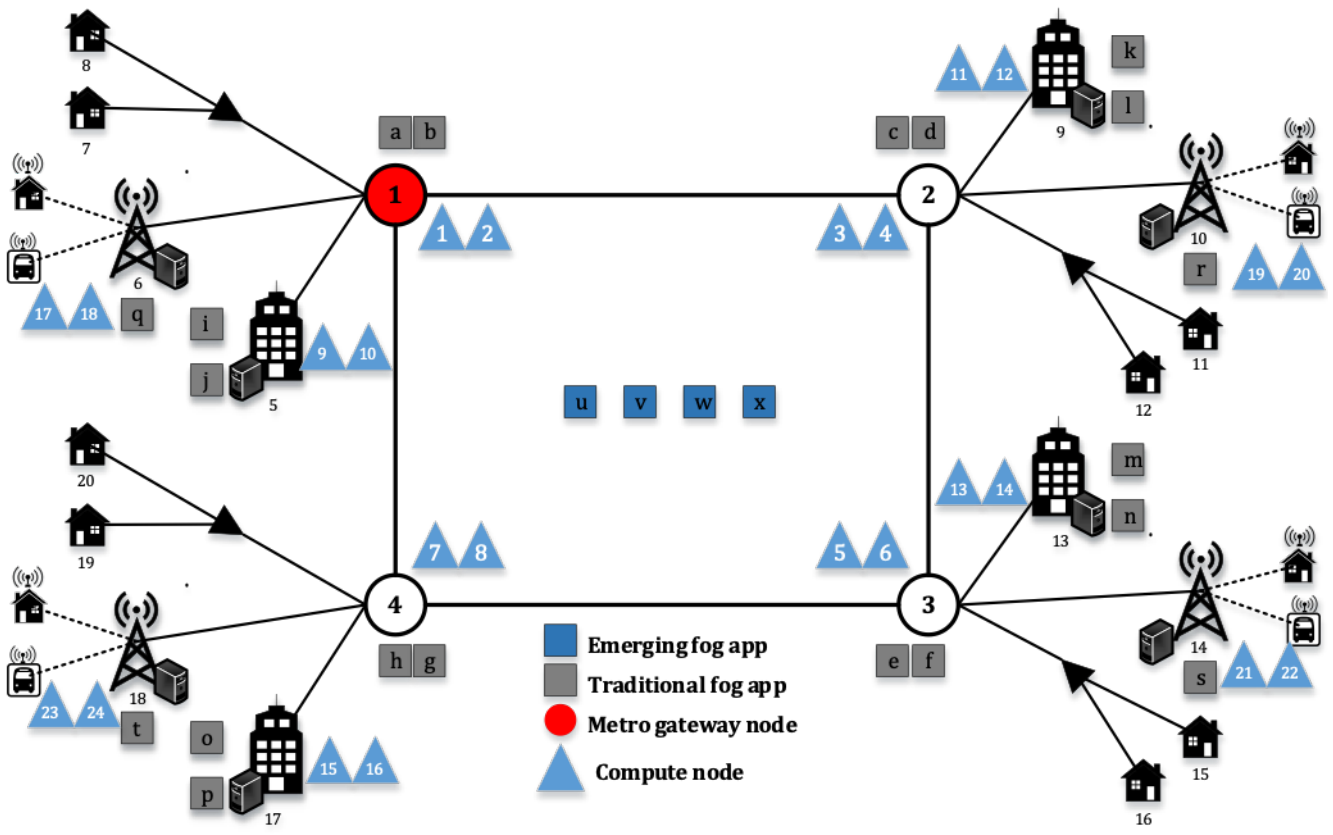

Fig. 2. Fog network system setup. The figure shows the evaluated network topology, the preferred fog computing sites, and the allocation of compute nodes to the fog sites. The figure also shows the traditional fog apps associated with each fog computing site and un-provisioned emerging fog apps.

Each designated fog computing site within the federation of fog computing nodes has one or two in-situ VM/VNFs for mission critical traditional applications (TA), as illustrated in Fig. 2, which must be provisioned at the node. Each CO node has 2 VNFs; each EO has 2 VMs; and each radio CS has 1 VNF. The resource demand of each mission critical TA is illustrated in Table II. Fig. 3a and Fig. 3b show the corresponding utilization of each compute resource components across distributed fog nodes under TS and DS architectures respectively after optimal placement of mission critical TAs. These figures show the presence of unused computing capacity to support emerging fog applications in preferred fog computing sites after TAs have been provisioned under both TS and DS architectures. Relative to the TS architecture, the DS architecture has greater average active resource utilization and reduced number of active resource components across fog computing sites. It is expected that these advantages of DS architecture over TS architecture will be maintained when attempts are made to place emerging fog apps in parallel with mission critical TAs. However, to achieve optimal efficiency and minimal rejection of emerging fog apps, the placement of each TA within each network node may be revised according to the characteristics of the emerging fog apps under consideration. Furthermore, the analysis of application placement is focused on emerging fog applications alone since the placement of mission critical TAs is fixed to specific fog computing sites.

TABLE I
COMPONENT CAPACITY AND POWER FEATURES

| Parameter | Value |
|---|---|
| CPU capacity, CPU peak power, and idle fraction of peak CPU power | 3.6GHz, 130W and 0.7 |
| Memory capacity, memory peak power, and idle fraction of peak memory power | 32GB, 11.85W and 0.7 |
| Storage capacity, storage peak power, and idle fraction of peak storage power | 320 GB, 6.19W and 0.7 |
| Metro Ethernet CPE (on-off) | 75 W [42] |
| Metro Ethernet aggregation router (load proportional) | 0.9 W/Gb [43] |
| Metro Ethernet access router (load proportional) | 0.243 W/Gb |
| PON optical line terminal (load proportional) | 1.75 W/Gb |
| PON optical network unit (on-off) | 15 W [44] |

Four types of emerging fog applications, which have distributed users in the access layer, are considered. Based on the assumption that all applications required by each enterprise are either hosted locally as VMs or hosted remotely in centralized hyper-scale DCs, distributed users of emerging fog applications are not associated with EOs. All user demands for a fog app in each access node are grouped together to form a cluster of user demand in that node. In all scenarios, a group of 5 end-users, which are attached to each radio CS via wireless media, collectively form a single clustered demand for each emerging fog app at the access layer. While a single end-user located in each residential house, forms a single clustered demand for each emerging fog app in the access layer.

Both VM/VNF of traditional fog apps and emerging fog apps have a mix of resource intensity as illustrated in Table II. Relative to the maximum compute resource capacity some apps are CPU intensive while others are memory intensive. Relative to the capacity of computes resource adopted, fog app "U" has medium CPU resource demand, high memory resources demand and low storage resource demand.  Fog app "V" has high CPU demand, high memory demand and medium storage demand, Fog app "W" has medium CPU demand, low memory, and low storage demand. Fog app "X" has low CPU, memory, and storage demands. Relative to other emerging fog apps, fog apps U and X have the highest pre and post processing traffic per user as illustrated in Table II.

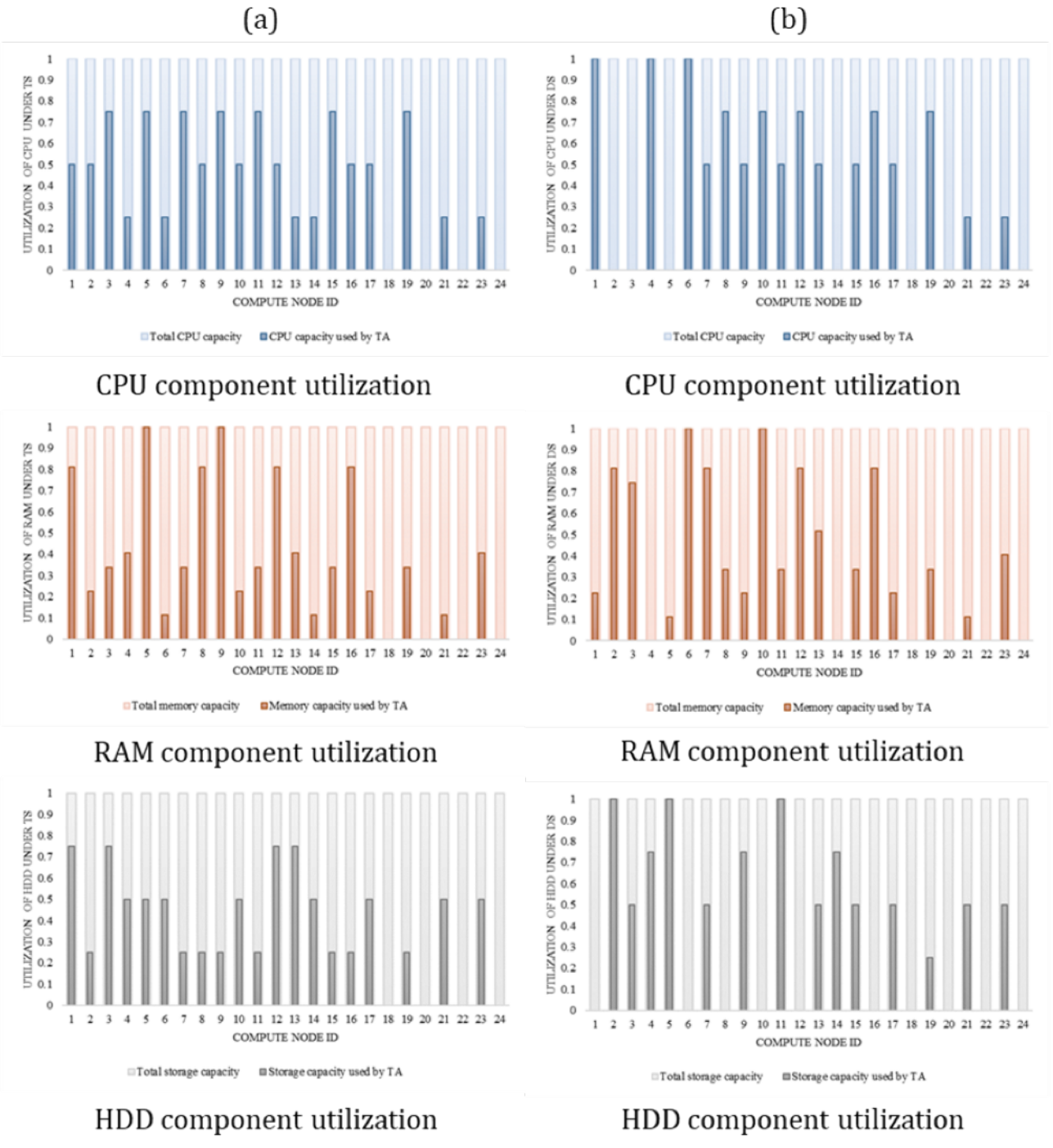

Fig. 3. (a) Resource component utilization under TS architecture. (b) Resource component utilization under DS architecture.

Although the placement of apps in centralized cloud computing DCs is not explicitly considered, the impact of cloud destined traffic on the overall performance of metro and access network tiers is considered and it is further assumed that traffic to and from the centralize cloud DC is part of the regular traffic traversing the network topology. A scenario where priority is given to traffic of emerging fog apps in the fog network is considered while traffic of other applications and services contribute to the regular traffic traversing metro and access networks. The range of regular traffic on the metro ring and access links are 114 – 120 Gbps and 4 – 5 Gbps respectively. Therefore, regular traffic utilizes about 60% and 12.5% of the capacity of a metro and access link respectively. The pre-processing and post-processing data rates per user for each emerging fog app considered are given in Table II. After processing at the optimal location, processed data is sent to the requesting users at the edge of the network and to the central cloud for further/historical analysis and persistent storage. It is assumed that the ratio of post-processing data rate to pre-processing data rate is about 50% as illustrated in Table II. In the worst-case scenario, if the request made by a given user (located at access node 7, a residential house with a single user for each emerging fog app) for all emerging fog apps is satisfied about 6% of the access link capacity will be utilized. On the other hand, if the requests made by all users in node 6 (a radio CS with multiple users for each emerging fog app) are satisfied, about 30% of the access link capacity will be utilized. Generally, compared to the fog related traffic, regular traffic is dominant in the network topology.

A scenario where shared network elements such as Ethernet access and aggregation routers and optical line terminals (OLTs) are assumed to have a load proportional power profile is considered. On the other hand, dedicated network components such as consumer premises equipment (CPE) and optical network units (ONUs) have an on-off power profile. Table I shows the power profile of each network element and their corresponding values. The exponential M/M/1 delay graph of each network link is divided into 6 linear pieces to implement piecewise linearization of the non-linear delay function. Fig. 4 gives the piecewise linear approximation of both 200 Gbps and 40 Gbps network links using the 6 linear pieces. The predefined upper bound for link load on both 200 Gbps and 40 Gbps network links are 195 Gbps and 39 Gbps respectively. Both values enforce a corresponding upper bound for queuing delay on each link. These values maybe varied based on expected network performance as desired by the network service provider on the corresponding link.

TABLE II
RESOURCE DEMAND OF FOG APPLICATIONS

| Fog App | CPU demand (GHz) | Memory demand (GB) | Storage demand (GB) | Pre-processing data rate per user (Gbps) | Post-processing data rate per user (Gbps) |
|---|---|---|---|---|---|
| A | 1.8 | 7.2 | 80 | - | - |
| B | 1.8 | 26 | 240 | - | - |
| C | 2.7 | 10.8 | 240 | - | - |
| D | 0.9 | 13 | 160 | - | - |
| E | 0.9 | 3.6 | 160 | - | - |
| F | 2.7 | 32 | 160 | - | - |
| G | 1.8 | 26 | 80 | - | - |
| H | 2.7 | 10.8 | 80 | - | - |
| I | 2.7 | 32 | 80 | - | - |
| J | 1.8 | 7.2 | 160 | - | - |
| K | 1.8 | 26 | 240 | - | - |
| L | 2.7 | 10.8 | 80 | - | - |
| M | 0.9 | 3.6 | 160 | - | - |
| N | 0.9 | 13 | 240 | - | - |
| O | 2.7 | 10.8 | 80 | - | - |
| P | 1.8 | 26 | 80 | - | - |
| Q | 1.8 | 7.2 | 160 | - | - |
| R | 2.7 | 10.8 | 80 | - | - |
| S | 0.9 | 3.6 | 160 | - | - |
| T | 0.9 | 13 | 160 | - | - |
| U | 1.8 | 26 | 120 | 0.9 | 0.45 |
| V | 2.7 | 32 | 160 | 0.11 | 0.05 |
| W | 1.8 | 7.2 | 80 | 0.83 | 0.41 |
| X | 0.9 | 3.6 | 40 | 0.43 | 0.22 |

This study evaluates the energy efficient placement of delay sensitive emerging fog applications in the presence of mission critical traditional fog applications in a shared distributed fog network. The MILP model is solved using the 64-bit AMPL/CPLEX solver on the ARC3 supercomputing node with 24 CPU cores and 128 GB of memory [41]. Analysis of results from the model focuses on metrics such as TFPC, TNPC, number of fog app instances created, roundtrip delay experienced by users of emerging fog applications. The number of active resource components across all fog computing sites in fog network and the corresponding average utilization of each active component type across the fog network is also adopted as an evaluation metric. To obtain optimal results, the results show that the MILP model effectively bin-packs workloads resource demands onto fog computing resources to achieve optimal resource power and utilization efficiencies within capacity constraints and limited resource utilization scope.

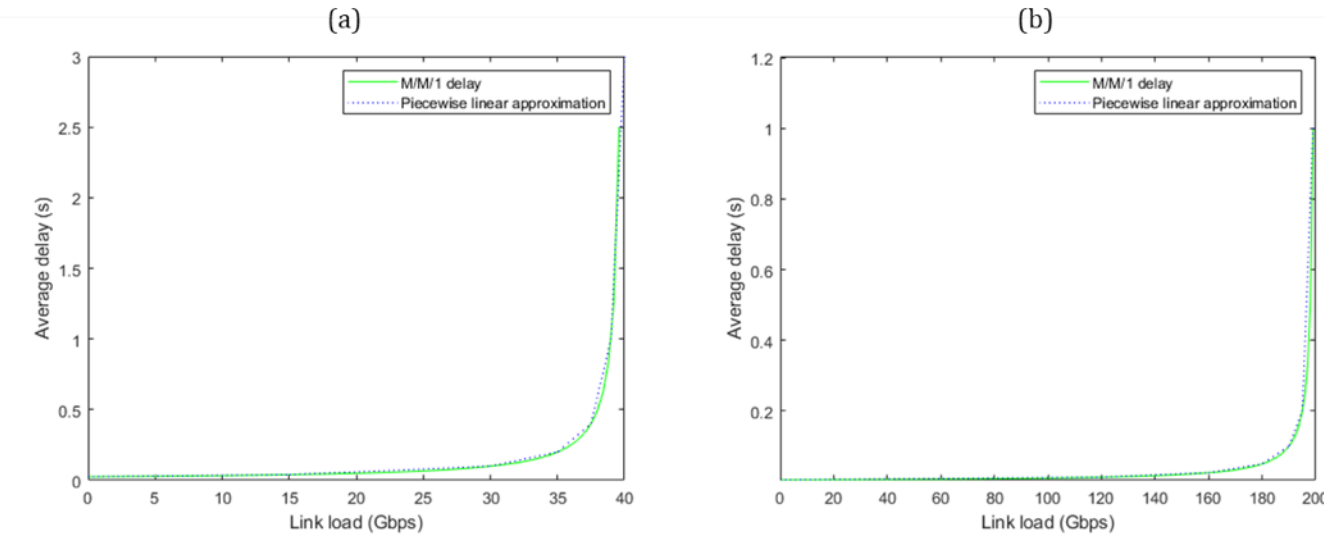

Fig. 4. (a) M/M/1 average delay of 40 Gbps access link. (b) M/M/1 average delay of 200 Gbps metro ring link.

### *B. Energy Efficient Placement under Low Delay Penalty*

Under this scenario, VMs/VNFs of both mission critical TAs and emerging fog applications are optimally placed in the fog network. All emerging fog apps considered are moderately sensitive to end-to-end delay in the network and the network has a trivial queuing delay penalty i.e. $\delta \ll 1$. The scenario is evaluated when TS architecture is adopted and when the DS architecture is adopted in fog computing servers.

#### *1) Placement under Traditional Server Architecture*

The illustration in Fig. 5 shows the optimal placement of emerging fog applications when TSs are deployed in the fog network. A single instance of each emerging fog app is provisioned in the fog network. The instance provisioned for each emerging fog app satisfies the computing capacity requested by all geo-distributed users of that app. These provisioned instances are strategically placed in the fog network to minimize both TNPC and TFPC. All mission critical TAs are also placed in their respective associated nodes to minimize the high cost of rejecting them as defined in the objective. However, network power consumption is dominant because a small fog network with a limited number of fog computing nodes, users and applications is considered. Moreover, fixed regular traffic in the network topology accounts for a significant portion of the TNPC.

Given enough CPU, memory and storage resource capacity in metro COs, the placement of fog apps as illustrated in Fig. 5 shows that there is high preference for metro COs when the selection of optimal locations for emerging fog apps is made. Relative to other network nodes, COs are centrally located, closer to geo-distributed users and closer to the metro gateway to the cloud. Hence, placement of fog apps in COs reduces the number of hops travelled by the fog traffic and therefore reduces TNPC. As illustrated in Fig. 5, instances of three emerging fog apps are placed in metro COs. Because homogenous servers are adopted across the distributed fog computing nodes, the power consumption incurred by hosting a given emerging fog app in inactive resource components is the same across all network nodes. Likewise, power consumption incurred by hosting a given emerging fog app in the idle resource capacity (IRC) of active resource components is also equal across all network nodes. Hence, multiple candidate metro COs which lead to the same increment of TFPC may exist for a given emerging fog app. Such ties are broken by selecting the metro CO which enables lower total (i.e. fog computing plus network) power consumption. Network node 1, which is also the metro gateway node to the core network in the network topology, wins such tie breaks. This is because the placement of emerging fog apps close to the metro gateway helps to reduce the number of hops traversed in the network topology and consequently the TNPC. Emerging fog apps "W" and "X" are placed in network node 1 as shown in Fig. 5. However, when resource capacity in network node 1 is limited, other candidate metro COs with adequate resource capacity are considered. Consequently, network node 3 is selected to host emerging fog app "U" as shown in Fig. 5.

In the absence of enough resource capacity in fog computing nodes attached to metro COs, fog sites in the access network must be selected to support emerging fog apps in the fog network if such fog sites have adequate computing capacity. However, multiple candidate access nodes may also present equal compute energy efficiency to host a given emerging fog application (as result of the homogeneous power profile of fog computing nodes across the distributed fog network). Hence, network energy efficiency is used as a decision metric to select the optimal network node in such scenarios. For example, emerging fog app "V" which requires a dedicated server because of the intensive nature of its memory resource demand is placed in network node 6, an access node as illustrated in Fig. 5. This node is selected over other network nodes (10, 14 and 18) due to its proximity to the metro gateway node which promotes lower total network traffic because the number of hops traversed by cloud bound traffic is reduced. It is important to note that unused servers are also present in network nodes 10, 14 and 18 as illustrated in Fig. 3a.

| Scenario | Low delay penalty | | | | High delay penalty | | Low delay penalty with delay sensitive fog app "X" | | | |
|---|---|---|---|---|---|---|---|---|---|---|
| Server Architecture | TS | | DS | | TS | DS | TS | | DS | |
| Network Node (type) | MILP | HEEDAP | MILP | HEEDAP | MILP | | MILP | HEEDAP | MILP | HEEDAP |
| 1 (CO) | ◆★ | ◆★ | | | ◆★ | | ◆ | ◆ | | |
| 2 (CO) | | | | | ◆ | ◆● | | | | ✚ |
| 3 (CO) | ● | ● | | | ● | | ● | ● | | |
| 4 (CO) | | | ◆★ | ◆★ | | ◆★ | | | | ◆ |
| 5 (EO) | | | | | | | | | | |
| 6 (Radio CS) | ✚ | ✚ | ● | | ★ | ● | ★✚ | ★✚ | ★✚◆ | ★ |
| 7 (Home) | | | | | | | | | | |
| 8 (Home) | | | | | | | | | | |
| 9 (EO) | | | | | | | | | | |
| 10 (Radio CS) | | | ★ | | ★ | ★ | ★ | ★ | ★ | ★ |
| 11 (Home) | | | | | | | | | | |
| 12 (Home) | | | | | | | | | | |
| 13 (EO) | | | | | | | | | | |
| 14 (Radio CS) | | | | ● | ◆★ | ✚ | ★ | ★ | ★ | ★ |
| 15 (Home) | | | | | | | | | | |
| 16 (Home) | | | | | | | | | | |
| 17 (EO) | | | | | | | | | | |
| 18 (Radio CS) | | | ✚ | ✚ | ◆★✚ | ●★ | ★ | ★ | ★● | ★● |
| 19 (Home) | | | | | | | | | | |
| 20 (Home) | | | | | | | | | | |

● - Fog app U ✚ - Fog app U ◆ - Fog app W ★ - Fog app X

Fig. 5. Energy efficient placement of emerging fog apps in a fog network.

#### *2) Placement under Disaggregated Server Architecture*

Replacing TSs with DSs in the fog network leads to changes in the optimal placement of emerging fog applications as shown in Fig. 5. Improved consolidation of both traditional and emerging fog apps enabled by the adoption of DSs in the distributed fog network is responsible for the revised placement observed. Consequently, Fig. 6d shows corresponding increases in the average utilization of active resources components in the fog network as result of the revision in server architecture. A primary instance of each emerging fog app is provisioned in the network node that leads to optimal energy efficiency when the DS architecture is deployed in the fog network. Furthermore, additional instances of an emerging fog app are created if such instances lead to marginal rise in TFPC while achieving to a significant drop in the TNPC. Reductions in TNPC is achieved because the creation of additional instance(s) enable reductions in the number of hops traversed. This explains the creation of two instances of emerging fog app “X” where the instance in network node 18 is responsible for distributed users of the application in network nodes 6, 7, 8, 14, 15, 16, 18, 19 and 20. A second instance of emerging fog app “X” in network node 10, which is provisioned using IRC of active resource components, is responsible for distributed users in network nodes 10, 11, and 12. Hence, the number of hops between instances of app “X” and their distributed users is minimized.

The revisions in fog apps placement observed when DSs are deployed is responsible for the fall (about 18%) in TFPC relative to results obtained when TSs are deployed in the fog network as shown in Fig. 6a. Reduction in the TCPC is responsible for over 90% of the fall seen in TFPC. This is because power consumption of CPU resource component is significantly higher than that of memory and storage resource components. Disaggregation enables significant improvements in CPU utilization efficiency (as shown in Fig. 6d) via improved consolidation of CPU resource demands of mission critical traditional apps and emerging fog apps in each fog computing site. Hence, the number of active CPU component reduced when DS are deployed to replace TS in the fog network as shown in Fig. 6c. The same is also true for HDD resource components and their corresponding utilization efficiency. However, the 33% drop in the number of HDD resource component observed in Fig. 6c does not lead to significant fall in the TFPC because HDD resource components have a lower peak power consumption relative to CPU and memory resource components as illustrated in Table I. Fig. 6c only shows a marginal drop in the number of active memory resource components. This is because several considered applications have high memory resource demand relative to the capacity of the homogenous memory resource components as given in Table II. Hence, a significant improvement in active memory resource utilization could not be realized as shown in Fig. 6d.

Fig. 6a shows a marginal increase in the TNPC after TSs are replaced with DSs in the fog network. This marginal rise is as a result of increased hop count between the instance of some emerging fog apps and their users. Furthermore, the hop count between instance of emerging fog app “W” and “X” and the metro gateway node also increases after a change in server architecture. Hence, network traffic traverses through more network equipment when DSs are deployed in the fog network. Fig. 6b gives the average and maximum round trip time between distributed user of each emerging fog app and the provisioned instances of the app. Relative to the deployment of TS architecture in the fog network, both average and maximum RTT increased when DS architecture is adopted in the fog network. Users of emerging fog app “U” experience relatively higher delay as shown in Fig. 6b compared to other emerging fog apps. Although the placement of fog app “U” in network node 6 enables optimal energy efficiency in the fog network, this choice also leads to high congestion on the link that connects network node 6 to node 1. Consequently, users of emerging fog apps “V”, “W” and “X”, which are located in node 6, experience the corresponding maximum delay illustrated in Fig. 6b. Since, the moderate delay thresholds of all emerging fog apps under this scenario are satisfied, such performance is acceptable. However, the performance obtained under both TS and DS architectures violates the delay threshold (i.e. $\tau \leq 20ms$) for delay sensitive fog apps. Hence, another subsection considers a scenario where requests for a delay sensitive emerging fog app are present in the fog network.

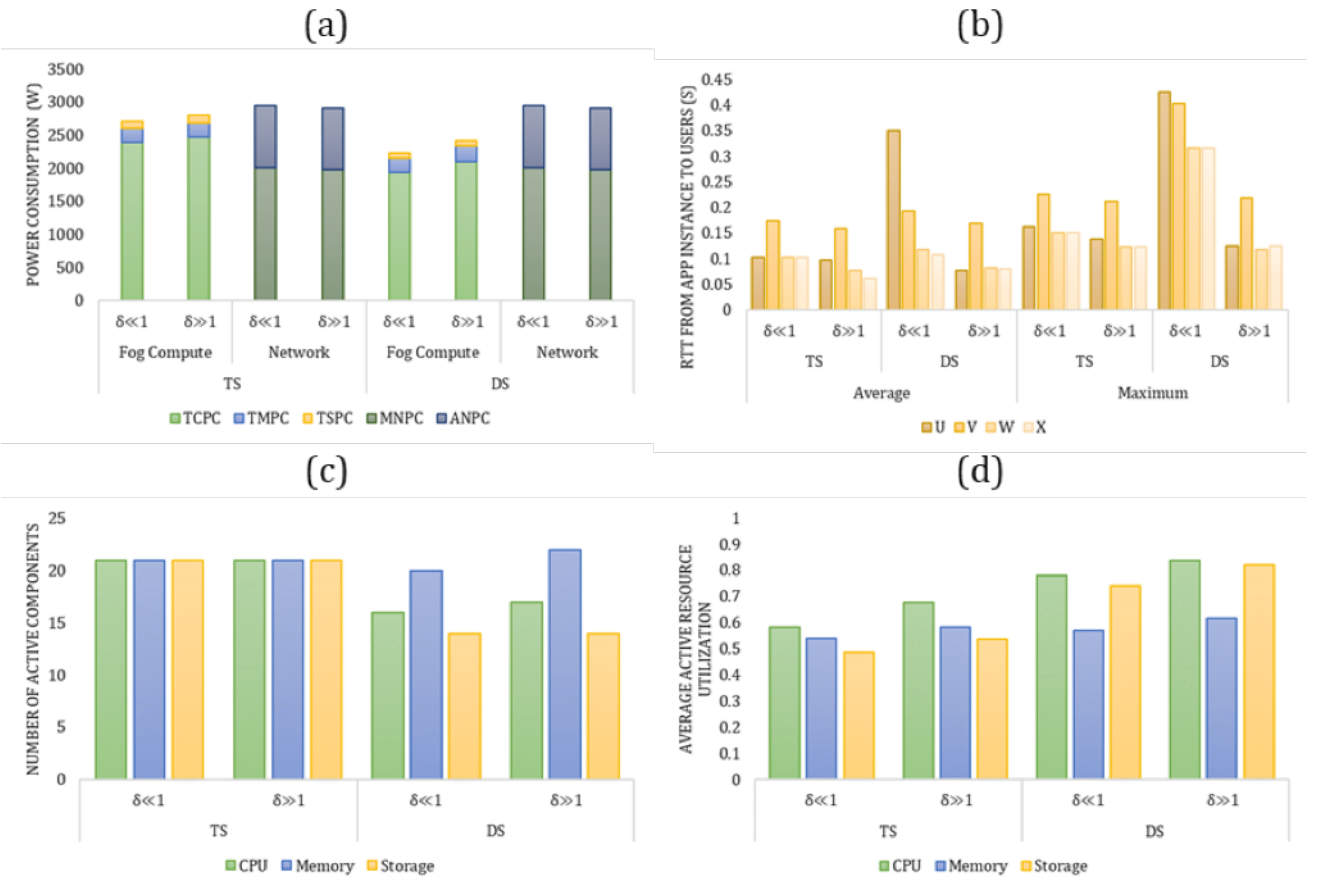

Fig. 6. (a) Power consumption under EE placement scenarios. (b) Round trip time from app instance to users under EE placement scenarios. (c) Number of active components under EE placement scenarios. (d) Average utilization of active resource components across fog computing sites under EE placement scenarios.

### C. *Energy Efficient Placement under High Delay Penalty*

In this subsection, $\delta \gg 1$; hence, network delay penalty is high. This represents a network where the operator desires minimal impact of emerging fog apps on regular traffic.

#### 1) *Placement under Traditional Server Architecture*

Under this scenario, multiple instances of some emerging fog apps are created when the TS architecture is adopted as shown in Fig. 5. This reduces the number of hops between instances of replicated fog apps and their users. Consequently, the total volume of traffic traversing the network topology is reduced and the delay experienced on each link of the network topology is also minimized. Relative to result obtained under low delay penalty ($\delta \ll 1$), the average and maximum RTT between the instance of fog apps and their distributed user falls when $\delta \gg 1$ as shown in Fig. 6b. However, to ensure a balanced trade-off between minimizing TFPC and the total approximated delay, only applications (app “W” and app “X”) with low-medium resource demand intensity are replicated as shown in Fig. 5. The primary instances for both apps are placed in centrally located fog sites (i.e. COs). These primary instances are responsible for users in directly attached access network nodes and also for most distributed users in the network topology. On the other hand, additional instances of emerging fog app “W” and app “X” are provisioned in some radio CSs. This strategy reduces the traffic associated with densely populated user clusters attached to each radio CS in the network and consequently reduces network congestion. Because the app "W" and app "X" have small compute footprint, they are easily provisioned with IRC of active TSs. Hence, minimal increase in TFPC is incurred compared to the $\delta \ll 1$ scenario as shown in Fig. 6a. Furthermore, relative to results obtained when $\delta \ll 1$, additional compute components are not activated to provision the additional instances of app "W" and "X" as shown in Fig. 6c. However, replication of fog app instances leads to the increase observed in the utilization of active resource components in the fog network as shown in Fig, 6d. Relative to the low delay penalty scenario where TSs are deployed, there is a 3% rise in the TFPC when the high delay penalty scenario is considered under a similar setup. On the other hand, the TNPC consumption fell by 2% as shown in Fig. 6a.

*2) Placement under Disaggregated Server Architecture*

A similar trend is observed when DSs are deployed in the fog network. Replicas of certain emerging fog apps are made, as shown in Fig. 5. This strategy enables reductions in the volume of traffic traversing the network topology and increased network congestions that may arise due to the creation of a single instance of each fog app. Consequently, distributed users of fog apps experienced lower average and maximum round trip time to assigned fog instances as shown in Fig. 6a. Under the high delay penalty scenario, three instances of app "U" and app "X" are created while only two instances of fog app "W" are created. High CPU and memory resource demands of emerging fog app "V" prevents the replication of the app to avoid significant increase in TFPC. As observed when TSs are deployed in the network, radio CSs are often used to host instances of emerging fog apps to ensure that the total volume of traffic in the network topology is minimized. This because of the high user density associated with radio CSs in the system setup. For example, App "V" is placed in network node 14 because the node does not require the activation of an additional CPU component to support the CPU intensive resource demands of emerging fog app "V".

Furthermore, since app "U" and app "W" have higher data rate per user relative to other emerging fog apps as illustrated in Table II, replication of these emerging fog apps can significantly reduce the total volume of traffic and congestion experienced in the network. Additionally, app "X" has non-intensive CPU and memory resource demands; therefore, the replication of app "X" does not lead to significant increase in the TFPC but can lead to additional reduction in the network traffic. Relative to the results obtained when low delay penalty is considered under a similar setup, the creation of replicas of some emerging fog apps when the DSs are deployed under the high delay scenario leads to about 8% rise in the TFPC as seen in Fig. 6a. A comparison of TNPC under both scenarios shows about 1% decrease due to reduced traffic on the network topology as illustrated in Fig. 6a. Under DS architecture, the creation of replicas of some emerging fog apps ("U", "W" and "X") under the high delay penalty increased the number of active compute resource components and the average active utilization of resource components in the fog network relative to the low delay penalty scenario as shown in Fig. 6c and Fig. 6d respectively.

Comparison of TS and DS architectures under the high delay penalty scenario expectedly shows that the adoption of the disaggregation concept enabled notable (about 14%) reduction in TFPC as shown in Fig. 6a. This is because server resource components are independently and proportionally utilized. A marginal fall in TNPC is also observed as a result of the revised server architecture. This is because the DS architecture encouraged the creation of more distributed replicas of most emerging fog apps relative to when the TS architecture is adopted in compute nodes within the fog network. Compared to the placement obtained under TS architecture where replication of emerging fog app "U" is discouraged because of the app's high compute footprint, replicas of emerging fog app "U" are created when the DS architecture is deployed. Note that app "U" has moderate CPU resource demand and high memory resource demand; hence, proportional usage of resource components when DS architecture is employed promotes the independent activation of new memory resource components to support replicas of app "U", while the moderate CPU resource demands of the app's replicas are aggregated with the CPU demand of other applications into active CPU components. However, replication of app "V" is still discouraged because of its high CPU and memory resource demands. Therefore, proportional usage of resource components does not enable sufficient benefits to promote replication of an emerging fog app that is CPU and memory intensive. Relative to the deployment of TS under the high delay penalty scenario, Fig. 6b shows that the average and the maximum round trip time are higher for some emerging fog apps when DSs are employed. Thus, the number of hops between users of such emerging fog apps and the instances of the app that is assigned to them increased because more network nodes are traversed.

### *D. Energy Efficient Placement of Delay Sensitive Fog App*

A network with trivial queuing delay penalty ($\delta \ll 1$) is adopted under this scenario. In contrast with the previous scenario, emerging fog app "X" is sensitive to end to end delay in the network (i.e. $\tau \leq 20ms$) under this scenario. Other emerging fog apps remain moderately sensitive to end-to-end delay in the network as in the previous subsections. Under this scenario, EE placement of both traditional and emerging fog apps is also evaluated when both TS and DS architectures are deployed in the fog computing nodes placed in the fog network.

Multiple instances of the delay sensitive emerging fog app are provisioned at all radio CS in the network topology as shown in Fig. 5 irrespective of the server architecture adopted in fog nodes. This ensures that the delay threshold of the fog app "X" is satisfied for users that are directly attached to a radio CS. On the other hand, users of emerging fog app "X" which do not have direct access to a radio CS are out rightly rejected. Hence, local computation capacity, which can lead to higher CAPEX and OPEX, is required to support such users. It is expected that a similar placement strategy will be implement if users of emerging fog apps are associated with enterprise office. Other emerging fog apps, which are moderately sensitive to delay, are placed in the fog network as shown in Fig. 5 to achieve optimal energy efficiency in the fog network as reported in Section

IV.B. Fig. 7b shows that the TNPC increases marginally when DS architecture is deployed in fog sites to replace TS architecture as observed previously.

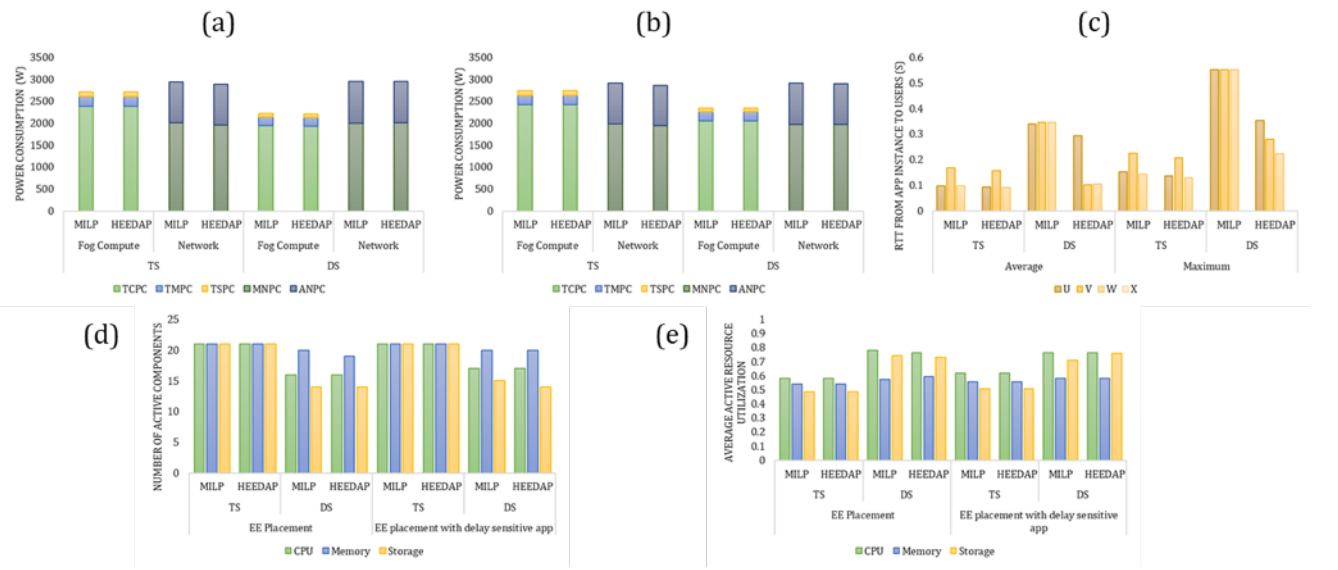

Fig. 7. (a) Power consumption under energy efficient placement scenario (b) Power consumption under energy efficient placement of a delay sensitive fog app scenario (c) Round trip time experienced by users of emerging fog app. (d) Number of active resource components in the fog network. (e) Average utilization of active resource components across fog computing sites in the network.

Furthermore, the TFPC and the number of active components is lower when DSs are used to replace TSs in the distributed fog network as shown in Fig. 7b and Fig. 7d respectively. By reducing the number of active resource components while provisioning the same number of emerging fog apps as when TS architecture is adopted, the adoption of DS architecture in the fog network can increase the amount of spare capacity available to support both highly sensitivity and moderately sensitive emerging fog apps at the network edge without additional CAPEX. As observed in Section IV.B, Fig. 7c shows that the average and maximum delay experienced by distributed users of emerging fog apps, which are moderately sensitive to end to end delay, increases when DS architecture is adopted. For instance, the illustration in Fig. 5 shows that each instance of moderately sensitive emerging fog apps created are provisioned in radio CSs, which are far from most of the distributed users of each fog app. However, the performance of such applications does not degrade because they have greater delay-tolerance. It is important to note that the delay experienced by served users of the delay sensitive apps is considered to be extremely low and insignificant as expected of today’s 5G mobile networks and future 6G mobile networks.

## V. Heuristic for Energy Efficient and Delay Aware Placement Of Fog Applications

The results obtained in the previous section reveal the ability of the formulated MILP model to perform best-effort placement of mission critical and emerging fog applications in a fog network in an energy efficient manner without violating delay requirement. However, due to the complexity of the MILP model, the performance of the MILP model is only evaluated for a small fog network. This complexity increases exponentially as the size of the fog network is scaled-up. Hence, massive computing capacity and time (days) are required to solve the MILP model for practical deployment in large fog networks. This is neither energy efficient nor practical because control and orchestration mechanisms of fog networks must make placement decisions in near real-time to ensure optimal user experience and fog applications’ performance. Therefore, a fast heuristic that mimics the insights obtained from the MILP model is a more ideal solution for fog networks.

To obtain results that approach those of the MILP model, a policy that leverages on a centralized orchestration and management framework for a network of distributed fog computing nodes is required. Such centralized control is an essential tool that can enable the efficiencies observed during the analysis of results from the solved MILP model. Hence, a heuristic is proposed; the heuristic depends on centralized control of distributed fog nodes and on global knowledge derived from control information exchange to achieve high efficiency in a fog network. The heuristic is a real-time algorithm that optimally provisions both mission critical traditional applications and (delay sensitive) emerging fog applications in a fog network when possible. The algorithm is derived from insights obtained from previous section.

Given a set of input fog apps (i.e. mission critical traditional apps and emerging fog apps), the algorithm attempts to provision instance(s) of these applications in a fog network in an energy efficient manner while considering the delay requirements of each application. Applications that cannot be provisioned are rejected and users whose delay requirements are not satisfied by provisioned instance(s) are also rejected. The algorithm supports the use of TS and DS architectures in the fog network. Other inputs to the algorithm include the user distribution and delay requirement of each emerging fog app; the distribution of fog compute nodes in the network topology; the features and characteristics of each resource components at each fog site; and the load and propagation delay on each network link.

### A. *HEEDAP Algorithm Description*

A high-level description of the heuristic for energy efficient and delay aware placement (HEEDAP) of applications in fog networks is illustrated in Fig. 8. At inception, the HEEDAP algorithm processes the set of input mission critical TAs which are associated with specific network nodes by sorting the list of TAs (in each network node) in descending order of CPU demand intensity. If a tie occurs, memory demand intensity is initially adopted to break the tie followed by storage demand intensity. The output of this process is the “local job list” (LJ-list) created at each network node.

Secondly, the HEEDAP algorithm processes the set of input emerging fog apps to the fog network in two stages. The initial stage identifies fog apps that are highly sensitive to network delay. These fog apps form the secondary job list at each fog computing site if some users of such delay sensitive fog app are directly connected to the corresponding network node via wireless media. This secondary job list created at each fog computing site is called the “pseudo-local job list” (PLJ-list). Emerging fog apps that are classified as delay sensitive are subsequently ejected from the list of input emerging fog apps. The PLJ-list at each network node is also arranged in descending order of resource demand intensity as described for the LJ-list. In the second stage, the HEEDAP algorithm sorts the list of input (moderately sensitive) emerging fog apps in descending order of resource intensity to create the “real fog job list” (RFJ-list). The RFJ-list also (implicitly) holds information about the number of users at each network node which is a source of any request made for each emerging fog app. After input apps processing, HEEDAP creates a temporary copy of the RFJ-list. This temporary copy is the "pseudo fog job list" (PFJ-list) which is refreshed after each complete iteration of the algorithm. It is important to note that an iteration of the HEEDAP algorithm is complete when the PFJ-list of that iteration is empty. Furthermore, a union of LJ-list and PLJ-list at each network node and the RFJ-list form the list of applications in the fog network.

Whilst the list of applications in the fog network is not empty (this is the first check of the HEEDAP algorithm), at each network node with compute capacity, the mission critical traditional fog app at the top of the LJ-list is set as a "query app". The query app at each network node is placed energy-efficiently; new resource components may be activated to support the query app as required. If the query app could not be placed, it is rejected and removed from the LJ-list. The state of all compute resource components in each network node is recorded and stored by the central orchestrator. Thereafter, at each network node, a candidate app in the LJ-list is identified to maximize the utilization of the IRC of active components in this network node where possible. If a candidate app is not found in the LJ-list, the PLJ-list is checked for a candidate app. The search for a candidate local or pseudo-local app in each network node gives higher priority to maximum utilization of the IRC of active CPU resource components because CPU components consume more power than memory and storage components as illustrated in Table I. When the DS architecture is adopted, inactive memory and storage resource components within the same network node may be used to complement available IRC of CPU resource component. A similar approach may be adopted when the TS architecture is deployed and a single compute node has multiple intrinsic CPU, memory, and storage resource components; otherwise, the resource locality constraint of TS architecture is enforced.

If a candidate app is not found in the LJ-list or PLJ-list of a network node, the PFJ-list is searched to identify a moderately sensitive emerging fog app that can maximize utilization of the IRC of active resource components of such network node. This search gives priority to the emerging fog app with more stringent delay requirement if the node is within the round-trip delay threshold of one or more unserved users of that fog app. The search conducted at each network node provides control data that supports the placement of emerging fog apps in the PFJ-list in subsequent steps of the HEEDAP algorithm. Such control information provides the global knowledge required by the algorithm.

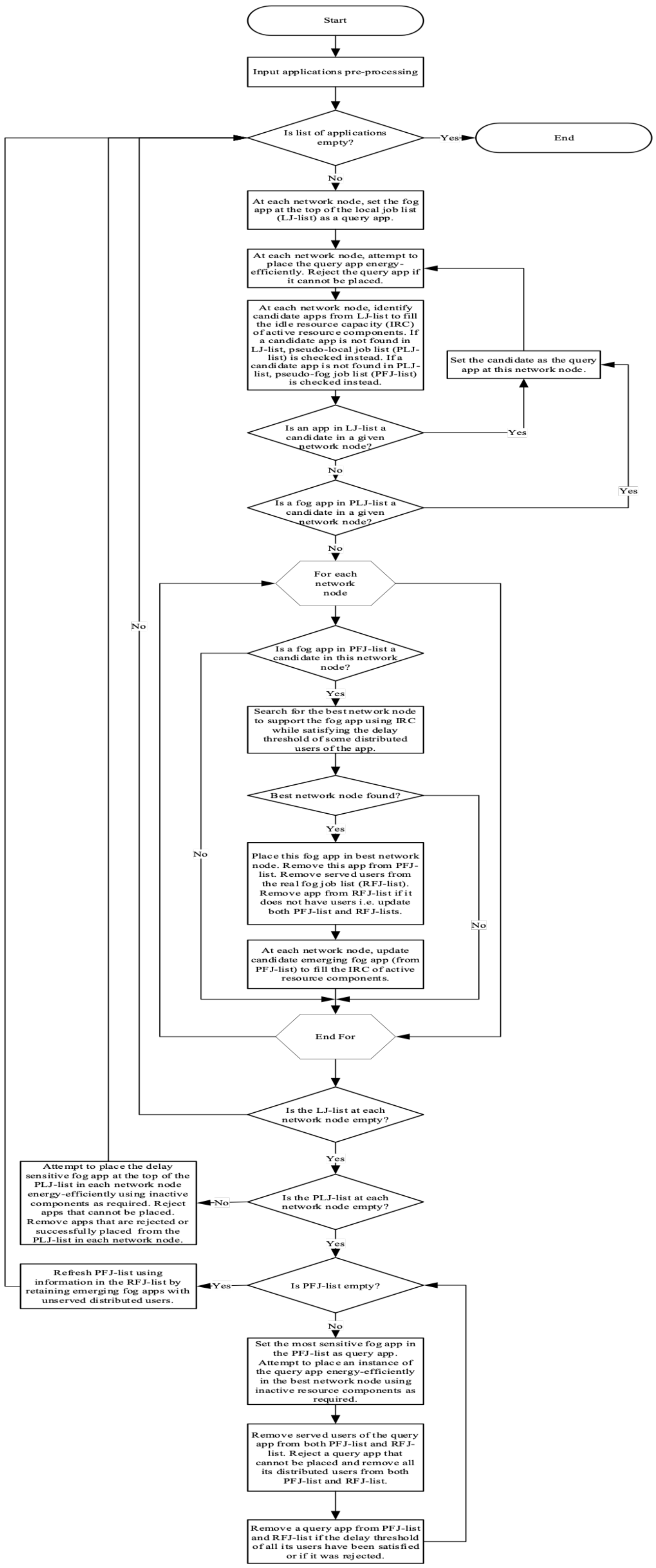

Fig. 8. Flow chart of HEEDAP algorithm.

In each network node, if a candidate app is found in the LJ-list, the app is selected as the new query app and placed energy-efficiently. Hence, the algorithm gives higher priority to mission critical traditional fog apps of the fog computing infrastructure provider. Otherwise, if a candidate app is found in the PLJ-list, the app is also selected as the new query app and is placed energy-efficiently. Relative to moderately sensitive emerging fog apps, this gives greater priority to emerging fog apps that have greater delay sensitivity. On the other hand, if a candidate app is not found in LJ-list or PLJ-list in a given network node and one is found in the PFJ-list, the algorithm checks if other network nodes can also host this candidate emerging fog app using IRC before a best network node is selected. Thus, global knowledge of the central orchestrator must be consulted before the best network node is selected from the list of all contending network nodes that can host the candidate moderately sensitive emerging fog app using IRC.

The best network node for a given moderately emerging fog app in the PFJ-List is the network node that leads to the smallest increase in TFPC after the placement of the candidate emerging fog app i.e. the most energy efficient network node. However, ties may occur when energy efficiency is used as the decision metric. Hence, the CO that is closest to the metro gateway node is given the highest priority when a tie occurs. This ensures that the (number of hops traversed or) traffic in the network is minimized as observed in the analysis of the MILP model results. Furthermore, if the list of contending network nodes for the emerging fog app comprise of only access network nodes, the network node with the highest user density is given higher priority. Otherwise, if some contending network nodes have the same user density for the emerging fog app, then lower delay to the metro gateway node is used as the decision metric to select the best network node.

Once the best network node for a given emerging in the PFJ-list fog app is found, an instance of the app is provisioned in that network node and all users of the app that can be served by this new instance (without violating delay requirements and network capacity constraints) are removed from the RFJ-list. Furthermore, the fog app and its served users are also removed from the PFJ-list of the present iteration. Additionally, the provisioned emerging fog app is replaced as a candidate emerging fog app at all corresponding fog computing sites where it was previously a candidate by electing a new candidate fog app from the PFJ-list to maximize the utilization of the IRC of active CPU component at such network nodes. This approach ensures fair placement for each emerging fog app in the fog network. This placement strategy, which fills the IRC of active CPU components in the fog network with emerging fog apps in the PFJ-list, is repeated at all network nodes to place all elected candidate emerging fog apps in the active iteration.

After exhausting all opportunities to use the IRC to provision an instance of each candidate emerging fog app in the PFJ-list in the active iteration, the size of the LJ-list is checked. If the LJ-list is not empty, the HEEDAP algorithm repeats all procedures described above; hence, the algorithm attempts to provision all mission critical traditional apps before considering emerging fog apps for placement. On the other hand, if the LJ-list is empty, another check is made to confirm that the PLJ-list in each network node is empty. If the PLJ-list in a network node is not empty, the HEEDAP algorithm attempts to provision the fog apps at the top of the PLJ-list in each network node. Such attempts may activate inactive resource components as required since the LJ-list in the network node is now empty. Hence, after mission critical TAs, emerging fog apps with higher sensitivity to delay have the highest priority. Emerging fog apps in the PLJ-list of a network node that are placed successfully are removed from the PLJ-list of that network node. Fog apps that are rejected are also removed from the PLJ-list of the corresponding network node. The HEEDAP algorithm thereafter repeats all procedures described above to place all moderately-sensitive emerging fog apps with available IRC of active resource components. It is important to note that if both LJ-list and PLJ-list of a network node are empty, only fog apps in the PFJ-list of the active iteration will be considered for energy efficient placement using IRC as given in the previous procedures described above.

On the other hand, if the PLJ-list is empty, a new check is made to confirm if the PFJ-list of the active iteration of the fog network is empty or not. If the PFJ-list is not empty, the emerging fog app with the highest delay sensitivity in the PFJ-list is set as query app and energy-efficient placement of the query app is attempted. Inactive resource components may be activated as required to place an instance of this query app. If a query app cannot be placed, the app is deleted from the PFJ-list and RFJ-list along with the information about all un-served distributed users of that app. Otherwise, if an instance of the query app was successfully provisioned, users of the query app whose delay threshold has been satisfied by the new instance are removed from both PFJ-list and RFJ-list. To ensure fairness when placing emerging fog apps, the provisioned query app is also removed from the PFJ-list of the active iteration, but the details of unserved distributed users are kept in the RFJ-list. Thereafter, the HEEDAP algorithm repeats all previous steps in this paragraph to place one instance of each moderately sensitive emerging fog apps in the PFJ-list until the list is emptied.

An empty PFJ-list implies that an active iteration of the HEEDAP algorithm has been completed and that a single instance of each emerging fog app has been provisioned. Since the HEEDAP algorithm uses the RFJ-list to maintain global knowledge of users of some emerging fog apps whose delay requirement remain unfulfilled, this knowledge is used to refresh PFJ-list. The refreshed PFJ-list comprise of all emerging fog apps with one or more unsatisfied users. The HEEDAP algorithm subsequently returns to the first check at the top of the algorithm to begin a new iteration since this check will be negative. However, if the delay requirements of all users of all emerging fog apps has been satisfied or the emerging fog apps has been rejected because their delay threshold could not be satisfied, the first check of the algorithm will be positive and the HEEDAP algorithm stops.

The HEEDAP algorithm calculates delay by considering the sum of the delay experienced on a link and link's propagation delay as the delay cost on each link in the network topology as given in the MILP model. The path with the smallest total delay is always selected as the shortest path between two nodes. It is assumed that the information of the network topology such as propagation delay and historical traffic (load) on each network link is available as input to the HEEDAP algorithm. Given this information, Dijkstra's shortest path algorithm is used determine the shortest path between two network nodes using total (propagation plus congestion) delay as the cost metric.

In the HEEDAP algorithm, resource locality constraint distinguishes a fog computing site with DS architecture from a fog computing site with TS architecture. To reduce the complexity of control and orchestration required for the algorithm in a large fog network deployment, big fog networks can be sub-divided into multiple small units. The algorithms can be deployed in a stand-alone mode in each small unit. Criteria for deciding the division thresholds for big fog networks include delay, number of network nodes and fog application user distribution. It is also important to note that an emerging fog app is a candidate app in a network node if and only if network capacity exists on a selected shortest path that satisfies the delay threshold of some users of that fog app after the placement of that fog app into the node. Otherwise, the fog app is not an acceptable candidate for that network node. Similarly, the users of placed fog app at a given network node are removed from the RFJ-list if and only if their delay threshold of such users are satisfied within specified link capacity constraint of links on the selected shortest path between users of that app and the network node where the instance has been provisioned.

## B. *HEEDAP Performance Evaluation*

To evaluate the performance of the HEEDAP algorithm, two evaluation scenarios (i.e. EE and EE placement of delay sensitive fog app) studied with the MILP model in previous section are considered. The results obtained when the algorithm is deployed in these scenarios are compared with those obtained by solving the MILP model. Similar computing and network metrics given in the previous section are also adopted. Since the evaluation of fog networks considered a network infrastructure that is shared with other services, it is assumed that fog network providers have SLAs with partners (network service providers) that guarantees shortest path delay for emerging fog apps in the network topology.

### 1) *Energy Efficient Placement*

In the absence of fog applications which are sensitive to delay under the EE placement scenario, Fig. 7a shows that the HEEDAP algorithm achieves comparable results as those reported when the MILP model is solved. As shown in Fig. 7, the HEEDAP algorithm achieves the same TFPC, number of active resource components and average active resource utilization as the MILP model when the TS architecture is deployed in the fog network. This demonstrates the efficacy of the HEEDAP algorithm at mimicking the compute energy efficiency achieved by the MILP model in a similar system setup that adopts homogenous resource components across the fog network. A single instance of each emerging fog app is provisioned in the fog network as reported when the MILP model was solved. As shown in Fig. 5 The placement of the instance created for each emerging fog app obtained via the HEEDAP algorithm is also an exact match with those obtained by solving the MILP model when TS architecture is deployed in fog nodes. Consequently, the average and maximum RTT from app instance to users obtained by HEEDAP are also comparable to those obtained by solving the MILP model when TS architecture is adopted in fog network nodes as shown in Fig. 7c. The TNPC obtained using the HEEDAP algorithm is marginally (about 2%) lower than that of the MILP model as shown in 7a. Disparity in path selection made for cloud bound traffic is responsible for this difference, while the MILP minimizes overall congestion in the network topology by distributing such traffic as necessary (which leads to higher network power consumption), HEEDAP always selects the shortest path which in turns minimizes network power consumption.

Similar trends are also observed when the HEEDAP algorithm is deployed to perform EE placement of moderately sensitive emerging fog apps in a fog network that adopts DS architecture. The TFPC obtained by HEEDAP algorithm is almost equal to that obtained by solving the MILP model. The TFPC of the HEEDAP algorithm is marginally (1%) lower because only a single instance of emerging fog app “X” is provisioned when HEEDAP algorithm is deployed as shown Fig. 5 while two instances of the same emerging fog app are created when the MILP model was solved. Consequently, relative to the TNPC obtained by solving the MILP model, the TNPC obtained by HEEDAP algorithm is marginally higher because the total volume of traffic in the network is higher. Furthermore, the placement of emerging fog apps obtained by the HEEDAP algorithm as shown in Fig. 5 is largely comparable to those obtained by solving the MILP model. However, compared to results obtained by solving the MILP model, the placement of emerging fog app “U” as obtained by the HEEDAP algorithm is different since the app is placed in node 14. This revised placement is responsible for the fall in the average and maximum RTT experienced by the distributed users of the app as shown in Fig. 7c. This is because node 14 is farther away from node 1 which is also the metro gateway node to the cloud. Hence, the congestion on the paths to node 14 is lower.

### 2) *Energy Efficient Placement of Delay Sensitive Fog App*

In the presence of a delay sensitive emerging fog application i.e. app X, the resulting placement of emerging fog apps, as depleted in Fig. 5, also shows the ability of HEEDAP algorithm to mimic the performance of the MILP when TS or DS architecture is adopted in the fog network. The pre-processing of input applications, which is performed in the initial steps of the HEEDAP algorithm, simplified the placement or rejection of delay sensitive fog apps in the presence or absence of in situ computing capacity at the source of each request for such apps. Therefore, the HEEDAP algorithm effectively mimicked the MILP model by provisioning some instances of delay sensitive emerging fog apps at radio CSs to serves users at such location. Users of delay sensitive emerging fog apps located at network nodes without local computing capacity are rejected by the fog network appropriately as discussed in the previous section.

When TS architecture is deployed in the fog network, the resulting placement of emerging fog apps by the HEEDAP algorithm is an exact replica of the placement obtained by solving the MILP model. Consequently, the same TFPC is achieved by both the MILP model and HEEDAP algorithm under the corresponding server architecture as shown in Fig. 7b.

As shown in Fig. 7d, the HEEDAP algorithm obtained the same number of active resource component as those obtained by solving the MILP model. Likewise, HEEDAP also replicates the average utilization of active components across fog computing sites obtained by solving the MILP model as show in Fig. 7e. The TNPC obtained by the HEEDAP algorithm is also comparable to the same value obtained by solving the MILP model under a similar scenario. Similarly, the average and maximum RTT to distributed users of emerging fog apps obtained by the MILP model is also comparable to those obtained by the solving the MILP model as shown in Fig. 7c.

However, when the DS architecture is employed in the fog network, the resulting placement of emerging fog apps by the HEEDAP algorithm is not an exact replica of the placement obtained by solving the MILP. The TFPC obtained with the HEEDAP algorithms is about 2% higher than the TFPC obtained by solving the MILP model under this scenario as shown in Fig. 7b. The adoption of homogeneous compute resource across the fog network is responsible for the comparable TFPC obtained. Therefore, the number of active resource components and the average utilization of these active components across fog computing sites as obtained by the HEEDAP algorithm is comparable to those obtained by solving the MILP model. Difference in the placement of emerging fog apps is responsible for the changes in the average and maximum RTT experienced by users as shown in **Error! Reference source not found.** 7c. Apps "V" and "W" are placed in COs in the metro ring by the HEEDAP algorithm; hence, the average and maximum RTT experience by the distributed users of these fog apps is reduced compared to results obtained by solving the MILP model under a similar setup.

## VI. Conclusions

In this paper, we evaluated the energy efficient placement of delay sensitive emerging fog applications in the presence of mission critical traditional fog applications in a shared distributed fog network that employs TS and DS architecture across selected fog computing sites at the network edge. Relative to the use of the TS architecture in the fog network that is built over a network with low delay penalty, the adoption of DSs enabled up to 18% reduction in TFPC. This because disaggregation enabled proportional usage of compute resources at each fog computing site and consequently improved the energy efficiency of the fog network. However, this is achieved at the expense of marginal increase in TNPC and somewhat higher response time when DS architecture is adopted. Setting up a fog network with high delay penalty increased the TFPC when either TS or DS architecture are employed in fog computing sites. This was done to minimize the congestion experienced on the network by reducing the network traffic. Consequently, the TNPC is also reduced. But, the TFPC of the fog network that employed DS architecture was 14% lower than that of the fog network that adopted TS architecture. Our result also showed that COs and radio CSs are important edge locations for supporting interactive applications demanded by geo-distributed end-users when energy efficiency is an important design criteria and such applications are moderately sensitive to the end-to-end delay experienced on the network. Otherwise, instances of such fog apps, which are more sensitive to delay, must be provisioned in the nearest network node that satisfies a predefined (and acceptable) delay threshold to distributed users. We also proposed a policy / heuristic, HEEDAP, which leverages on a centralized orchestration and management framework, for a network of distributed fog computing nodes. The policy is a real-time algorithm that optimally provisions both mission critical traditional applications and (delay sensitive) emerging fog applications in a fog network when possible. The HEEDAP algorithm achieves comparable results as those reported when the MILP model was solved under similar evaluation scenarios. On most occasions, the HEEDAP algorithm achieves the same application placement pattern, compute and network energy efficiencies as the exact results obtained by solving the MILP model. Occasional different between the result obtained via by the HEEDAP algorithm and by solving the MILP model are marginal. For example, the different between the TFPC obtained with the HEEDAP and that obtained by solving the MILP model is not greater than 2% in all scenarios considered. Future work will consider scenarios where mission critical traditional apps can be placed dynamically in contrast to the static placement considered in this paper. Furthermore, inter-workload communication may also be introduced between applications. The HEEDAP algorithm can also be extended to investigate various placement strategies for emerging fog apps. Finally, artificial intelligence and machine learning techniques can also be adopted to further validate and enhance the proposed concepts and policy in this paper.

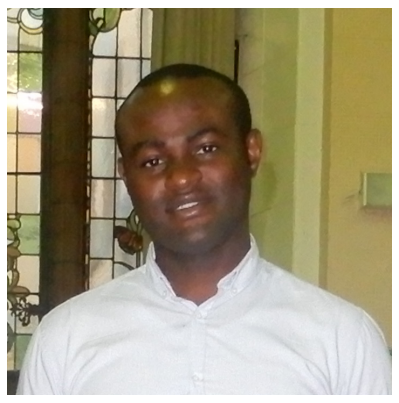

**Opeyemi O. Ajibola** received the B.Sc. degree (High Hons.) in Electrical and Electronic Engineering from Eastern Mediterranean University, Famagusta, North – Cyprus, in 2011 and the M.Sc. degree (with distinction) in Digital Communications Networks from University of Leeds, Leeds, UK in 2015. He is currently working towards the PhD degree in the School of Electronic and Electrical Engineering, University of Leeds, Leeds, UK. From 2012 to 2013, he was a Wireless Solution Sales Engineer with Huawei Technologies, Abuja, Nigeria. In 2014, he joined Federal University Oye-Ekiti, Ekiti State, Nigeria as a graduate assistant. His research interests include composable datacenter infrastructures, energy efficient datacenter and communication networks, energy efficient cloud and fog/edge computing and the Internet of Things.

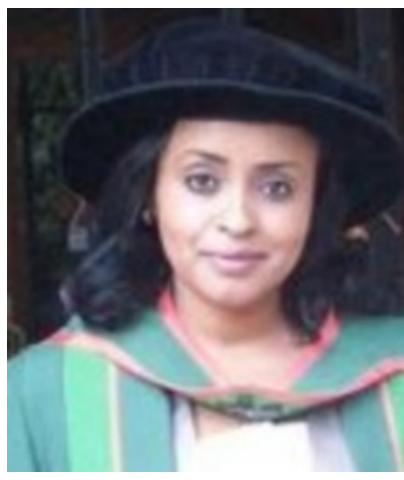

**Taisir E. H. El-Gorashi** received the B.S. degree (first-class Hons.) in Electrical and Electronic Engineering from the University of Khartoum, Khartoum, Sudan, in 2004, the M.Sc. degree (with distinction) in Photonic and Communication Systems from the University of Wales, Swansea, UK, in 2005, and the PhD degree in Optical Networking from the University of Leeds, Leeds, UK, in 2010. She is currently a Lecturer in optical networks in the School of Electronic and Electrical Engineering, University of Leeds. Previously, she held a Postdoctoral Research post at the University of Leeds (2010– 2014), where she focused on the energy efficiency of optical networks investigating the use of renewable energy in core networks, green IP over WDM networks with datacenters, energy efficient physical topology design, energy efficiency of content distribution networks, distributed cloud computing, network virtualization and big data. In 2012, she was a BT Research Fellow, where she developed energy efficient hybrid wireless-optical broadband access networks and explored the dynamics of TV viewing behavior and program popularity. The energy efficiency techniques developed during her postdoctoral research contributed 3 out of the 8 carefully chosen core network energy efficiency improvement measures recommended by the GreenTouch consortium for every operator network worldwide. Her work led to several invited talks at GreenTouch, Bell Labs, Optical Network Design and Modelling conference, Optical Fiber Communications conference, International Conference on Computer Communications, EU Future Internet Assembly, IEEE Sustainable ICT Summit and IEEE 5G World Forum and collaboration with Nokia and Huawei.

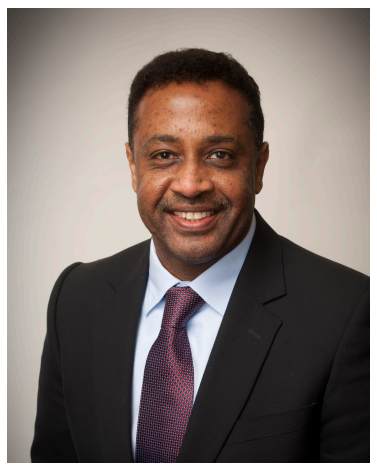

**Professor Jaafar Elmirghani** is Fellow of IEEE, Fellow of the IET, Fellow of the Institute of Physics and is the Director of the Institute of Communication and Power Networks and Professor of Communication Networks and Systems within the School of Electronic and Electrical Engineering, University of Leeds, UK. He joined Leeds in 2007 having been full professor and chair in Optical Communications at the University of Wales Swansea 2000-2007.

He received the BSc in Electrical Engineering, First Class Honours from the University of Khartoum in 1989 and was awarded all 4 prizes in the department for academic distinction. He received the PhD in the synchronization of optical systems and optical receiver design from the University of Huddersfield UK in 1994 and the DSc in Communication Systems and Networks from University of Leeds, UK, in 2012. He co-authored Photonic Switching Technology: Systems and Networks, (Wiley) and has published over 550 papers.

He was Chairman of the IEEE UK and RI Communications Chapter and was Chairman of IEEE Comsoc Transmission Access and Optical Systems Committee and Chairman of IEEE Comsoc Signal Processing and Communication Electronics (SPCE) Committee. He was a member of IEEE ComSoc Technical Activities Council' (TAC), is an editor of IEEE Communications Magazine and is and has been on the technical program committee of 41 IEEE ICC/GLOBECOM conferences between 1995 and 2020 including 19 times as Symposium Chair. He was founding Chair of the Advanced Signal Processing for Communication Symposium which started at IEEE GLOBECOM'99 and has continued since at every ICC and GLOBECOM. Prof. Elmirghani was also founding Chair of the first IEEE ICC/GLOBECOM optical symposium at GLOBECOM'00, the Future Photonic Network Technologies, Architectures and Protocols Symposium. He chaired this Symposium, which continues to date. He was the founding chair of the first Green Track at ICC/GLOBECOM at GLOBECOM 2011, and is Chair of the IEEE Sustainable ICT Initiative, a pan IEEE Societies Initiative responsible for Green ICT activities across IEEE, 2012-present. He has given over 90 invited and keynote talks over the past 15 years.

He received the IEEE Communications Society 2005 Hal Sobol award for exemplary service to meetings and conferences, the IEEE Communications Society 2005 Chapter Achievement award, the University of Wales Swansea inaugural 'Outstanding Research Achievement Award', 2006, the IEEE Communications Society Signal Processing and Communication Electronics outstanding service award, 2009, a best paper award at IEEE ICC'2013, the IEEE Comsoc Transmission Access and Optical Systems outstanding Service award 2015 in recognition of "Leadership and Contributions to the Area of Green Communications", the GreenTouch 1000x award in 2015 for "pioneering research contributions to the field of energy efficiency in telecommunications", the IET 2016 Premium Award for best paper in IET Optoelectronics, shared the 2016 Edison Award in the collective disruption category with a team of 6 from GreenTouch for their joint work on the GreenMeter, the IEEE Communications Society Transmission, Access and Optical Systems technical committee 2020 Outstanding Technical Achievement Award for outstanding contributions to the "energy efficiency of optical communication systems and networks". He was named among the top 2% of scientists in the world by citations in 2019 in Elsevier Scopus, Stanford University database which includes the top 2% of scientists in 22 scientific disciplines and 176 sub-domains. He was elected Fellow of IEEE for "Contributions to Energy-Efficient Communications," 2021.

He is currently an Area Editor of IEEE Journal on Selected Areas in Communications series on Machine Learning for Communications, an editor of IEEE Journal of Lightwave Technology, IET Optoelectronics and Journal of Optical Communications, and was editor of IEEE Communications Surveys and Tutorials and IEEE Journal on Selected Areas in Communications series on Green Communications and Networking. He was Co-Chair of the GreenTouch Wired, Core and Access Networks Working Group, an adviser to the Commonwealth Scholarship Commission, member of the Royal Society International Joint Projects Panel and member of the Engineering and Physical Sciences Research Council (EPSRC) College.

He has been awarded in excess of £30 million in grants to date from EPSRC, the EU and industry and has held prestigious fellowships funded by the Royal Society and by BT. He was an IEEE Comsoc Distinguished Lecturer 2013-2016. He was PI of the £6m EPSRC Intelligent Energy Aware Networks (INTERNET) Programme Grant, 2010-2016 and is currently PI of the EPSRC £6.6m Terabit Bidirectional Multi-user Optical Wireless System (TOWS) for 6G LiFi, 2019-2024. He leads a number of research projects and has research interests in communication networks, wireless and optical communication systems.